\documentclass[10pt,twoside]{article}
\usepackage[margin=1in]{geometry}
\usepackage{latexsym,amssymb,amsmath,amsfonts,amsthm,xcolor,cases,mathtools}
\usepackage{graphicx}
\usepackage{epstopdf,chngcntr}
\usepackage{natbib}
\usepackage{hyperref}

\numberwithin{equation}{section}
\theoremstyle{thmstyletwo}%
\newtheorem{remark}{Remark}%

\hypersetup{
    colorlinks = true,
    linkcolor = {blue},
    citecolor = {blue},
    urlcolor  = {blue},}

\newcommand{\RE}[1]{\textup{Re}\left\{#1\right\}}

\newcommand{\CC}{{\mathbb{C}}}

\newcommand{\ZZ}{{\mathbb{Z}}}

\newcommand{\ParDer}[2]{\frac{\partial#1}{\partial#2}}

\newcommand{\BF}[1]{\mathbf{#1}}

\definecolor{UoM}{RGB}{100,30,160}
\definecolor{newgreen}{RGB}{0,150,0}

\newcommand{\RED}{} % Without red comments
\usepackage[scr=boondox]{mathalfa}% pour mathscr
\newcommand{\gchoice}{\RED{\mathscr{C}}}

\begin{document}
\title{Acoustic wave diffraction by a quadrant of sound-soft scatterers$^1$}
% Acoustic wave diffraction by a doubly periodic quarter square lattice
\author{M. A. Nethercote$^2$,~A. V. Kisil$^3$~and~R. C. Assier$^3$}
\footnotetext[1]{Accepted by The IMA Journal of Applied Mathematics on 08/07/2025.}
\footnotetext[2]{Department of Applied Mathematics and Theoretical Physics, University of Cambridge, United Kingdom,\newline
	(mn598@cam.ac.uk.)}
\footnotetext[3]{Department of Mathematics, Alan Turing Building, University of Manchester, United Kingdom,\newline
	(anastasia.kisil@manchester.ac.uk, raphael.assier@manchester.ac.uk).}

\date{} 
\maketitle

%This article studies diffraction of acoustic waves by a doubly periodic quarter square lattice which consists of simple cylindrical scatterers with Dirichlet boundary conditions. We will be using a method outlined in \citep{Kisil2023} and comparing with our previous but similar method \citep{MSIApaper}. Traditionally, the Wiener--Hopf technique has been successful in solving many canonical problems with sharp transitions in boundary conditions on a plane/plate including diffraction problems with continuum wedges. For wedges with discrete boundaries, we have recently published an article which studied and found solutions to the problem of diffraction of acoustic waves by a wedge consisting of point scatterers. The methodology derived in that article was also generalised to an arbitrary number of periodic semi-infinite arrays with arbitrary orientations. Subject to a suitable truncation, we can use this method to construct and simulate the effect of doubly periodic wedge lattices by periodically positioning the arrays. For the specific case with a doubly periodic quarter square lattice, we take a different approach in this article by only considering the nearest-neighbour interactions as part of the Wiener--Hopf kernel instead of decomposing the lattice into rows or columns. 

\begin{abstract}
Motivated by research in metamaterials, we consider the challenging problem of acoustic wave scattering by a doubly periodic quadrant of sound-soft scatterers arranged in a square formation, which we have dubbed the quarter lattice. This leads to a Wiener--Hopf equation in two complex variables with three unknown functions for which we can reduce and solve exactly using a new analytic method. After some suitable truncations, the resulting linear system is inverted using elementary matrix arithmetic and the solution can be numerically computed. This solution is also critically compared to a numerical least squares collocation approach and to our previous method where we decomposed the lattice into semi-infinite rows or columns.
\end{abstract}

\section{Introduction}
Multiple scattering of waves (scattering involving many obstacles) is a fast growing area of research due to increasing computational and theoretical capabilities~\citep{pmartin2006,McIver2007, HewettHewitt2016,CHCA2019,MSIApaper}. It is becoming of increased importance as more and more materials rely on structuring to obtain desirable properties. One example is the increased interest in metamaterials~\citep{SchnitzerCraster2017,Putley2023,CrasterGuenneau2023}. This has been facilitated by the rapid growth and availability of 3D printing, making previously infeasible experimental projects now achievable.

Analytical methods for multiple scattering problems typically rely on specific configuration of scatterers. For example, the Wiener--Hopf method is a versatile method \citep{LawrieAbrahams2007,KisilWHReview2021} that has been used to study semi--infinite arrays of point scatterers under Foldy's approximation~\citep{HillsKarp1965,LintonMartin2004}. This method has also been applied to the multiple scattering problem with a semi--infinite lattice in~\citep{TymisThompson2011,TymisThompson2014}, where the Wiener--Hopf technique was used on a system of equations reduced by periodicity. One feature of the semi-infinite geometry discussed in these articles is that within a stop band, the scattering coefficients will decay to zero with respect to how deep one is within the lattice and any Bloch waves that could occur come from zeros in the Wiener--Hopf kernel.

Recently, the authors have further extended this approach to work for any combination of semi--infinite arrays in a semi--analytic method \citep{HWpaperI,MSIApaper}. In this work, we aim to use an analytical approach based on the newly developed extension to the Wiener--Hopf method~\citep{Kisil2023} to consider a quadrant of small cylindrical Dirichlet scatterers \RED{of radius $a$} arranged in a square lattice formation, for which we can use Foldy's approximation \citep{Foldy1945} \RED{(see also \citet{ParnellAbrahams2010} where an interesting discussion of Foldy's approximation in multiple scenarios, including the sound-soft case, can be found)}. We will call these \textit{Foldy's point scatterers} and we refer to this problem as the \textit{quarter lattice diffraction problem}. This is an interesting scattering problem since there are elements of periodicity (within the quadrant the scatterers are arranged periodically) and there is a boundary in which the scatterers are positioned. It is known that both periodicity and boundaries significantly influence the scattering of waves. These two effects together make the problem both challenging and interesting to study.

More specifically, the scattering of a plane wave by a quadrant of Foldy's point scatterers (see Figure \ref{fig:QL-diagram} and \ref{fig:QL-highlighted}) can be reduced to the solution of the following equation for the scattering coefficients $A_{pq}$
\begin{align}\label{QL-SoEs}
A_{pq} \gchoice +\hspace{-1em} \sum_{\substack{m,n=0\\(m,n)\neq (p,q)}}^{\infty} \hspace{-1em} A_{mn}\mathbb{H}_{(p-m)(q-n)} 
=-e^{-iks\Theta(p,q)},\quad p,q\geq0,
\end{align}
where 
\begin{align}\label{QL-SoEs-hankel}
\gchoice&\RED{=H_0^{(1)}(ka),} \quad \mathbb{H}_{(p)(q)}=H^{(1)}_0\left(ks\sqrt{p^2+q^2}\right), \quad \RED{\Theta(p,q) =p\cos(\theta_{\textrm{I}})+q\sin(\theta_{\textrm{I}})}.
\end{align}
%\begin{align}\label{QL-SoEs-angle}
%\Theta(p,q) =p\cos(\theta_{\textrm{I}})+q\sin(\theta_{\textrm{I}}),
%\end{align}
\RED{As discussed in Remark \ref{rem:rem1}, note that the choice of $\gchoice$ is not unique, though it is expected to depend on the product of the wavenumber $k$ and the cylinders' radius $a$. Note further that $\mathbb{H}_{(p)(q)}$ depends on the product of the spacing $s$ and wavenumber $k$, while $\Theta(p,q)$ depends} on the angle of incidence $\theta_{\textrm{I}}$ of the plane wave. Finding an analytical solution of this equation is not currently possible to the authors' knowledge, however, \eqref{QL-SoEs} can be rearranged \RED{as follows} to simplify the problem\RED{:}
\begin{align}\label{QL-SoEs-MSIA}
A_{pq}\gchoice + \sum_{\substack{n=0\\n\neq q}}^{\infty} A_{pn}\mathbb{H}_{(0)(q-n)} 
=-e^{-iks\Theta(p,q)}-\hspace{-0.5em} \sum_{\substack{m,n=0\\m\neq p}}^{\infty} \hspace{-0.5em} A_{mn}\mathbb{H}_{(p-m)(q-n)} ,\quad p,q\geq0.
\end{align}
The above rearrangement corresponds to interpreting the quarter lattice as an infinite set of coupled semi-infinite rows (or columns if we replace $\mathbb{H}_{(0)(q-n)}$ with $\mathbb{H}_{(p-m)(0)}$ and edit the sums accordingly). This allows us to essentially reuse our previous work \citep{HWpaperI,MSIApaper}, which applies the Wiener--Hopf method to find a solution to \eqref{QL-SoEs-MSIA}. The result is an infinite linear system which has to be truncated before it can be inverted to obtain the numerical solution for the scattering coefficients. However, this approach works better when there is more separation between the arrays which means that there will be a weaker interaction between them (for example, a quadrant of scatterers where the vertical spacing is larger than the horizontal spacing and the arrays are rows). Because we have a square formation here, there will be a stronger interaction and hence, in this article we consider a different rearrangement of \eqref{QL-SoEs} which is given in \eqref{QL-AK-SoEs}. As we will see in this article, this new rearrangement can also be approached via a Wiener--Hopf method. The problem has two perpendicular boundaries and hence, can be solved using the new analytic method developed in~\citep{Kisil2023}. This method builds on the traditional Wiener--Hopf method but the key difference is that the initial step is performing the discrete Fourier transform in both perpendicular directions. This leads to a functional equation in two complex variables (instead of one) with three unknown functions (instead of two) which can then be solved by exploiting the symmetries. This method has already been used in the context of elastic wave scattering in a structured quadrant~\citep{Nieves_2d_24}.

General Wiener--Hopf equations in two complex variables have recently been studied in diffraction theory for the quarter--plane problem \citep{AssierAbrahamsSIAP21}, the right--angled penetrable wedge problem \citep{KunzAssierSIAP2022} and the discrete right--angle problem \citep{ShaninKorolkov2020,ShaninKorolkov2022}. In general, there are no known techniques to solve them exactly. Nevertheless, information about the singularities of the unknown spectral functions can be unveiled through methods of analytic continuation \citep{AssierShaninQJMAM2018} \citep{KunzAssierQJMAM2023}, and recent development on asymptotic evaluation of double Fourier--type integrals \citep{AssierShaninKorolkovQJMAM2022} can be used to extract important information on the far--field behaviour of the physical field \citep{AssierShaninKorolkovQJMAM2024} \citep{KunzAssierIMA2024}. A similar approach can also be used for wave propagation within periodic structures \citep{ShaninAssierKorolkovMakarov2024}. The Wiener--Hopf equation present in this article \eqref{QL-AK-WHE} has a special form in the sense that some of its unknown functions depend on one variable only, allowing for mathematical progress as highlighted in \citep{Kisil2023}. 

There are also some very competent numerical techniques that can be used to determine the scattering behaviour for an arbitrary configuration of scatterers. Two examples of these techniques include a T--matrix reduced order model \cite{HawkinsGanesh2017,tmatsolver,Tmatpaper} and a least squares collocation approach that was used in \cite{ChapmanHewettTrefethen2015} and \cite{HewettHewitt2016} to study the electrostatic and electromagnetic shielding by Faraday cages. These methods are very efficient at modelling the individual interactions between scatterers, but they cannot model an infinite number of scatterers. Thus they cannot model the behaviour of scattering coefficients infinitely far from the origin of an infinite periodic quarter lattice.
%due to the fact that it must have a finite number of scatterers.

The structure of the paper is as follows. Firstly, we formulate the quarter lattice diffraction problem in Section \ref{Sec:setup}. Then we derive a functional Wiener--Hopf equation in two variables with three unknown functions, before defining and using a manifold to reduce this to a simpler Wiener--Hopf equation with one variable and one unknown function in Section \ref{Sec:Derive}. The latter Wiener--Hopf equation is solved in Section \ref{Sec:Solve} and the scattering coefficients are then determined in Section \ref{Sec:Amn}. Lastly in Section \ref{Sec:Results}, we display some test cases using this new method and perform a fair comparison against the results obtained using our previous approach described in \citep{MSIApaper} as well as other numerical methods. We use the dispersion diagram for the equivalent infinite lattice (see e.g. \citep{KrynkinMcIver2009}, \cite{EllisBarnwell-thesis} for details) to select frequencies that are expected to be in the stop and pass bands of the quarter lattice and investigate the behaviour of the scattering coefficients to see if it is as expected. 

\section{Formulation of the quarter lattice diffraction problem}\label{Sec:setup}
\begin{figure}[ht]\centering
\includegraphics[width=0.6\textwidth]{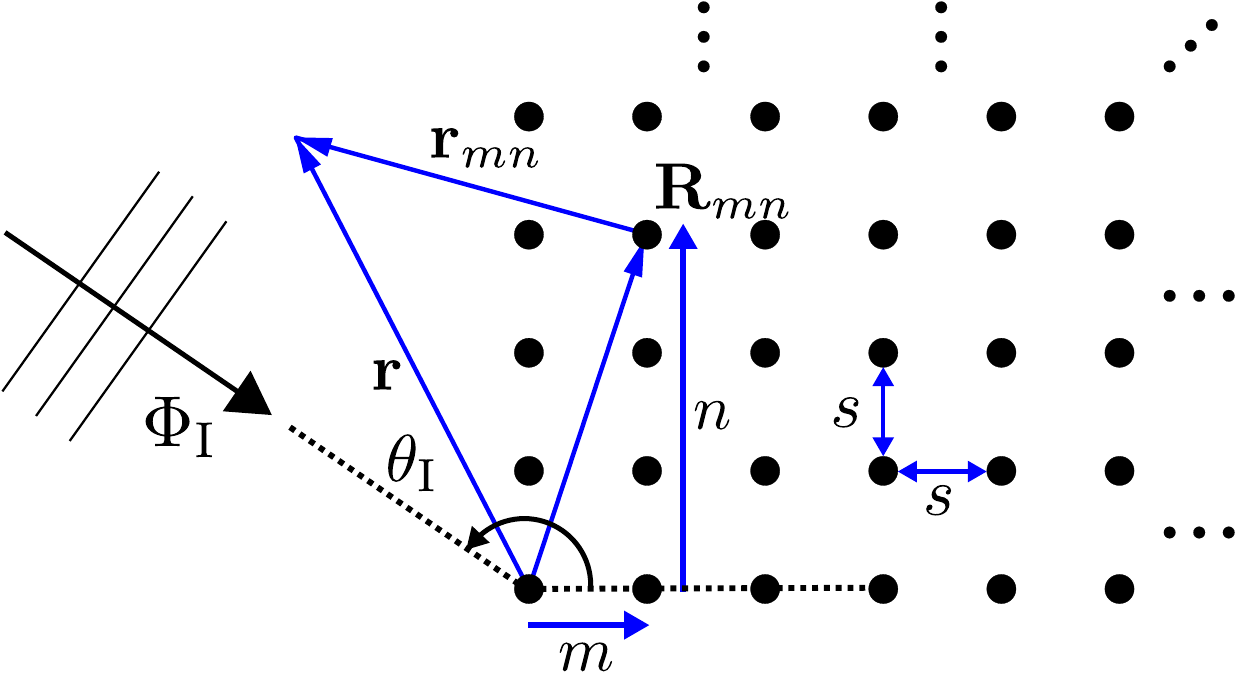}
\caption{Diagram of the quarter lattice problem where the scatterers are arranged in a doubly periodic and square formation with spacing given by $s$.}
\label{fig:QL-diagram}
\end{figure}

In the problem we are considering, the scatterers are arranged in a square formation, equally spaced with a horizontal and vertical spacing given by $s$. All the scatterers are simple circular cylinders with sound--soft (or homogeneous Dirichlet) boundary conditions and have the same radius $a$ which is assumed small. The scatterers are indexed by the discrete coordinates $(m,n)$ and the position of the centre of the scatterer at $(m,n)$ is given by,
\begin{align}\label{QL-R_mn}
\BF{R}_{mn}=ms\BF{\hat{x}}+ns\BF{\hat{y}},\quad m,n\geq0
\end{align}
where $(\BF{\hat{x}},\BF{\hat{y}})$ are the unit vectors of a Cartesian coordinate system centred at the corner of the quarter lattice. Similarly to \citep{MSIApaper}, we are looking for time--harmonic solutions to the linear wave equation by assuming and then suppressing the time factor $e^{-i\omega t}$, where $\omega$ is the angular frequency, and use a polar coordinate system $(r,\theta)$ with the position vector given by $\BF{r}$. We let $\Phi$ be the total wave field and decompose it into an incident wave field $\Phi_{\textrm{I}}$ and the resulting scattered field $\Phi_{\textrm{S}}$ by the equation $\Phi=\Phi_{\textrm{I}}+\Phi_{\textrm{S}}$. Each of these fields satisfy the Helmholtz equation with wavenumber $k$. The incident wave field $\Phi_{\textrm{I}}(\BF{r})$ takes the form of a unit amplitude plane wave given by,
\begin{align}\label{QL-PhiI}
\Phi_{\textrm{I}}=e^{-ikr\cos(\theta-\theta_{\textrm{I}})}
\end{align}
where $\theta_{\textrm{I}}$ is the incoming incident angle and $k=\frac{\omega}{c}$ is the wavenumber ($c$ being the speed of sound of the host medium). An important assumption of this study is the use of Foldy's approximation \citep{Foldy1945,pmartin2006}, where we assume that the cylinders are isotropic point scatterers. That requires them to be small in comparison to the wavelength (i.e.\ $ka\ll1$) and this allows us to write the scattered field $\Phi_{\textrm{S}}$ in the form of the monopole expansion
\begin{align}\label{QL-gensol}
\Phi_{\textrm{S}}(\BF{r})=\sum_{m,n=0}^{\infty} A_{mn}H^{(1)}_0(k|\BF{r}-\BF{R}_{mn}|),
\end{align}
where $A_{mn}$ are the scattering coefficients of the quarter lattice. Note that the subscript of the sum in \eqref{QL-gensol} is a shorthand for a double sum in $m$ and $n$ where $m$ and $n$ have the same range of values. A diagram of the problem we are considering is given in Figure \ref{fig:QL-diagram}.

To obtain the system of equations mentioned in the introduction, we use the same methods featured in \citep{LintonMartin2004} and \citep{HWpaperI}. This involves applying the Dirichlet boundary conditions at each scatterer and approximating for small scatterer size which leads to \eqref{QL-SoEs}. In the next section, we will rearrange and transform \eqref{QL-SoEs} into a Wiener--Hopf equation in two complex variables, and solve it using the methodology outlined in \cite{Kisil2023}.

\RED{
\begin{remark} \label{rem:rem1}
It is important to note that due to Foldy's approximation, the boundary conditions are only approximately satisfied. 
Regarding energy, \citet[Chap.~8]{pmartin2006} showed that it is conserved if
\begin{align}\label{QL-CoE}
\left|g_{mn}\right|^2+\RE{g_{mn}}=0,
\end{align}
where $g_{mn}=A_{mn}/\Phi_{mn}(\BF{R}_{mn})$ with $$\Phi_{mn}(\BF{r})=\Phi(\BF{r})-A_{mn}H^{(1)}_0(k|\BF{r}-\BF{R}_{mn}|).$$

In this work, to find $g_{mn}$, we follow \cite{HillsKarp1965} by first evaluating the above equation for $\BF{r}$ on the boundary of the $mn$ scatterer (where $\Phi$ is zero and $|\BF{r}-\BF{R}_{mn}|=a$). Then, using the facts that $\Phi_{mn}$ is well-behaved in the neighbourhood of $\BF{R}_{mn}$ and that $ka\ll1$, we approximate $\Phi_{mn}(\BF{r})$ by $\Phi_{mn}(\BF{R}_{mn})$. This leads to $g_{mn}=-(H^{(1)}_0(ka))^{-1}$ and implies our choice of $\gchoice=H_0^{(1)}(ka)$. For this choice of $g_{mn}$ (and $\gchoice$), the energy-conservation equation \eqref{QL-CoE} is not exactly satisfied, though it is asymptotically satisfied as $ka\to0$ since the quantity of interest is of the order $O\left(\left(ka/\ln(ka)\right)^2\right)$.

Although we do not follow this path in the present work, an alternative (but asymptotically consistent) choice of $g_{mn}$ would be to replace our choice by its leading asymptotic behaviour as $ka\to0$, i.e.\ taking $g_{mn}=-(\frac{2i}{\pi}\left(\ln(ka/2)+\gamma\right)+1)^{-1}$, where $\gamma$ is Euler's constant. Interestingly, this choice of $g_{mn}$, used for instance by \citet{LintonMartin2004}, conserves energy exactly in the sense of \eqref{QL-CoE}. 

Importantly, this alternative choice would not affect any of the mathematical developments provided in the rest of the article. The only change to make is to choose $\gchoice=\frac{2i}{\pi}\left(\ln(ka/2)+\gamma\right)+1$ instead of $\gchoice=H^{(1)}_0(ka)$.

%replace the term $H^{(1)}_0(ka)$ by $\frac{2i}{\pi}\left(\ln(ka/2)+\gamma\right)+1$ in \eqref{QL-SoEs} and wherever else $H^{(1)}_0(ka)$ appears in what follows. 

Note further that the essence of Foldy's approximation is to neglect multipole terms, which, in effect, results in neglecting terms of order $O((ka)^2)$ and smaller. Because the two possible choices of $g_{mn}$ mentioned above differ by $O\left((ka)^2/\ln(ka)\right)$, so will the associated computed wave fields. This discrepancy is a $o((ka)^2)$ and so both solutions will be equivalent asymptotically speaking.
We conclude this remark with the observation that another energy-conserving and asymptotically consistent choice could have been made by setting $\gchoice=H_0^{1}(ka)/J_0(ka)$.
%Note that the next multipole terms $H^{(1)}_1(z)$ (which were neglected when we assumed Foldy's approximation) will lead to a discrepancy of the order $O\left((ka)^2\right)$ and the two choices of $g_{mn}$ differ by $O\left((ka)^2/\log(ka)\right)$ which is smaller than this discrepancy. This means that any discrepancy between these two choice are negligible in our numerical results.
%$O\left((ka)^2\log(ka)\right)$, the final numerical results can only differ in the same way, which is what we observe indeed in our numerical experiments.
\end{remark}
%for $ka\ll1$, we approximate $\Phi_{mn}(\BF{R}_{mn})$ by $\Phi_{mn}$ on the boundary of the associated scatterer to obtain $g_{mn}=-(H^{(1)}_0(ka))^{-1}$.
% Following \cite{HillsKarp1965}, for $ka\ll1$, we approximate $\Phi_{mn}(\BF{R}_{mn})$ by $\Phi_{mn}$ on the boundary of the associated scatterer to obtain $g_{mn}=-(H^{(1)}_0(ka))^{-1}$. This leads to a discrepancy in \eqref{QL-CoE} of the order $O\Big(\left(\tfrac{ka}{\ln(ka)}\right)^2\Big)$, which is very small when $ka\ll1$.  Though we do not do this in the present work, it is possible for \eqref{QL-CoE} to be satisfied exactly by considering only the leading order approximation of $g_{mn}$ as did \citet{LintonMartin2004}, however this does not change the mathematics of this article (except for replacing $H^{(1)}_0(ka)$ with $\frac{2i}{\pi}\left(\ln(ka/2)+\gamma\right)+1$ in \eqref{QL-SoEs}) and the numerical results agree with each other at leading order with the difference being $O((ka)^2)$. 
} 
%This leads to 
%\begin{align}\label{QL-SoEs}
%A_{pq}H^{(1)}_0(ka)+\hspace{-1em} \sum_{\substack{m,n=0\\(m,n)\neq (p,q)}}^{\infty} \hspace{-1em} A_{mn}\mathbb{H}_{(p-m)(q-n)} 
%=-e^{-iks\Theta(p,q)},\quad p,q\geq0.
%\end{align}
%where $\mathbb{H}_{(p)(q)}=H^{(1)}_0\left(ks\sqrt{p^2+q^2}\right)$ and $\Theta(p,q)=p\cos(\theta_{\textrm{I}})+q\sin(\theta_{\textrm{I}})$. 
\section{Deriving and solving the Wiener--Hopf equation}\label{Sec:Derive}
\subsection{Derivation of the generalised Wiener--Hopf equation}
Firstly, we identify the nearest neighbours of a scatterer $(p,q)$ and define the set $S_{pq}=\{(m,n):\ m\in[p-1,p+1],\ n\in[q-1,q+1]\}\ \cap\ \ZZ_{\geq0}^2$ and the rest as $\bar{S}_{pq}=\ZZ_{\geq0}^2\backslash S_{pq}$ (see Figure \ref{fig:QL-highlighted}). Then we rearrange \eqref{QL-SoEs} moving all the terms not associated to the neighboring scatterers to the right hand side with the forcing terms, to obtain: 

\begin{figure}[ht]\centering
\includegraphics[width=0.6\textwidth]{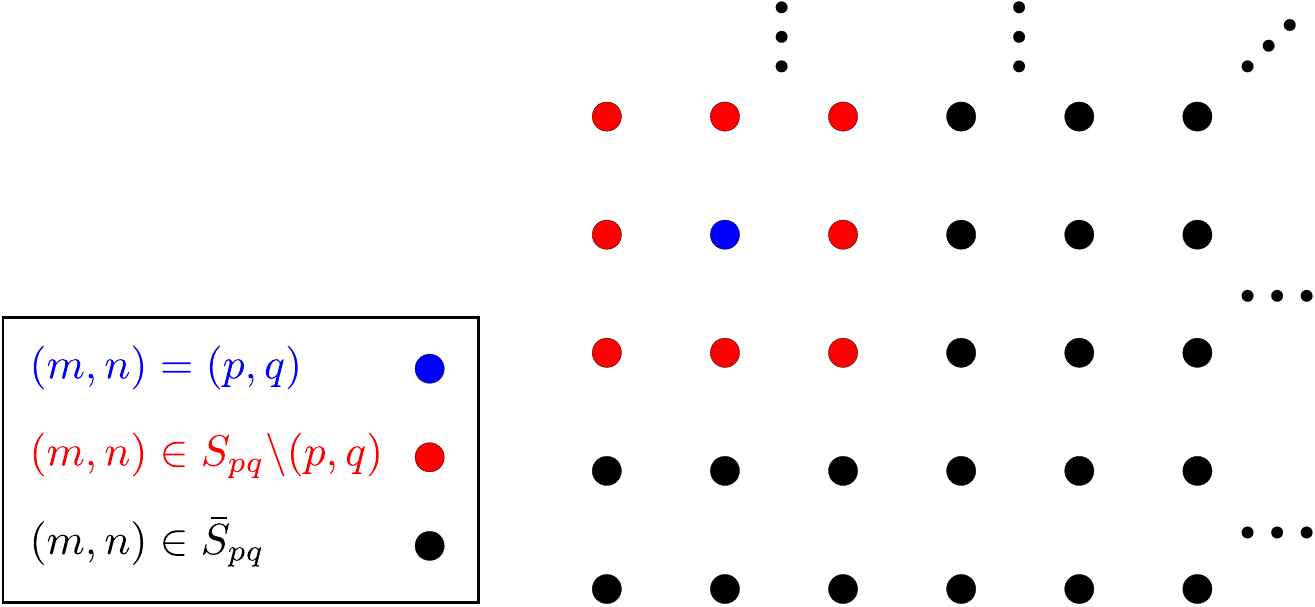}
\caption{Here we illustrate how we split the double sum in the system of equations given by \eqref{QL-AK-SoEs}. Say $A_{pq}$ is associated with the blue scatterer, then the coefficients associated with the neighbouring red scatterers stay on the left-hand side, whereas all other ones associated with the black scatterers move to the right-hand side.}
\label{fig:QL-highlighted}
\end{figure}
\begin{align}\label{QL-AK-SoEs}
A_{pq} \gchoice &+\hspace{-1em} \sum_{\substack{m,n\in S_{pq}\\(m,n)\neq (p,q)}}\hspace{-1em} A_{mn}\mathbb{H}_{(p-m)(q-n)}=-e^{-iks\Theta(p,q)}-\hspace{-1em} \sum_{(m,n)\in \bar{S}_{pq}}\hspace{-1em} A_{mn}\mathbb{H}_{(p-m)(q-n)},\quad p,q\geq0.
\end{align}
where, $\mathbb{H}_{(p-m)(q-n)}$ and $\Theta(p,q)$ are defined in \eqref{QL-SoEs-hankel}. Recall that $A_{pq}=0$ automatically for all $p,q<0$ because there are no scatterers present. We show in Appendix \ref{App:Func_eqn} that by multiplying \eqref{QL-AK-SoEs} by $z^{p+1}\zeta^{q+1}$ and summing for all $p,q\geq0$, we obtain the following generalised Wiener--Hopf equation in two complex variables:
\begin{align}\label{QL-AK-WHE}
z\zeta K(z,\zeta)A^{++}(z,\zeta)-zL_2(z)B^{+}_1(z)-\zeta L_2(\zeta)B^{+}_2(\zeta)+A_{00}H^{(1)}_0\left(ks\sqrt{2}\right) =z\zeta F^{++}(z,\zeta).
\end{align}
The three unknown functions $A^{++}(z,\zeta)$, $B^{+}_1(z)$ and $B^{+}_2(\zeta)$ are defined as the following double and single half range Z--transforms, 
\begin{align}\label{QL-AK-A++}
A^{++}(z,\zeta)=\sum_{p,q=0}^{\infty}A_{pq}z^p\zeta^q,\\
\label{QL-AK-B+}
B^{+}_1(z)=\sum_{p=0}^{\infty}A_{p0}z^p,\ \ B^{+}_2(\zeta)=\sum_{q=0}^{\infty}A_{0q}\zeta^q.
\end{align}
The known kernel function $K$ is defined by,
\begin{align}\label{QL-AK-kernelK}
K(z,\zeta)=& L_1(z)+\left(\zeta+\frac{1}{\zeta}\right)L_2(z)\\%=L_1(\zeta)+\left(z+\frac{1}{z}\right)L_2(\zeta),\\
%=&H^{(1)}_0(ka)+\left(z+\frac{1}{z}+\zeta+\frac{1}{\zeta}\right)H^{(1)}_0(ks) +\left(z+\frac{1}{z}\right)\left(\zeta+\frac{1}{\zeta}\right)H^{(1)}_0\left(ks\sqrt{2}\right),\\
\label{QL-AK-kernelL1}\textrm{where},\quad L_1(z)=& \gchoice +\left(z+\frac{1}{z}\right)H^{(1)}_0\left(ks\right),\\ 
\label{QL-AK-kernelL2}L_2(z)=&H^{(1)}_0(ks)+\left(z+\frac{1}{z}\right)H^{(1)}_0\left(ks\sqrt{2}\right),
\end{align}
and the forcing function $F^{++}$ (which includes both known and unknown parts) is defined by,
\begin{align}\label{QL-AK-WHE-forcing}
\nonumber F^{++}(z,\zeta)&=-\frac{1}{(1-z_cz)(1-z_s\zeta)}-\sum_{p,q=0}^{\infty}S_A(p,q)z^p\zeta^q,\\
%\nonumber F_{\text{etc}}(z,\zeta)&=\\
\textrm{where},\quad S_A(p,q)&=\sum_{(m,n)\in \bar{S}_{pq}}A_{mn}\mathbb{H}_{(p-m)(q-n)}, %H^{(1)}_0\left(ks\sqrt{(p-m)^2+(q-n)^2}\right),
\end{align}
$z_c=e^{-iks\cos(\theta_{\textrm{I}})}$ and $z_s=e^{-iks\sin(\theta_{\textrm{I}})}$. Once we are able to solve this Wiener--Hopf equation, and hence find $A^{++}(z,\zeta)$, then the scattering coefficients can be recovered by the inverse transform given by 
\begin{align}\label{QL-Z-trans-inv}
A_{mn}=\frac{1}{(2\pi i)^2}\int_{C_z}\int_{C_\zeta}A^{++}(z,\zeta)z^{-m-1}\zeta^{-n-1}\textrm{d}\zeta\textrm{d}z,
\end{align}
where $C_z$ and $C_\zeta$ are unit circle contours in their respective complex domains. Note that the kernel $K$ is similar to the one given in \citep{ShaninKorolkov2020,ShaninKorolkov2022, Nieves_2d_24}. The difference in those articles is that, only the nearest horizontal and vertical interactions are considered whereas here we include the nearest diagonal interactions as well. 

Let us assume that the wavenumber $k$ has a small positive imaginary part. Then the Wiener--Hopf equation is valid in the region $(z,\zeta)\in \Omega^+_z\times \Omega^+_\zeta$ where the two regions,
\begin{align}\label{QL-AK-plus-regions}
\Omega^+_z=\{z:|z|\leq1/|z_c|, z\neq 1/z_c\},\quad \Omega^+_\zeta=\{\zeta:|\zeta|\leq1/|z_s|, \zeta\neq 1/z_s\},
\end{align}
both contain the unit circle in their respective complex planes. It is important to note that any functions with a $+$ superscript are analytic in the region $\Omega^+_\bullet$ in the associated complex plane (and similarly $++$ superscripts are associated with functions analytic in $\Omega^+_z\times \Omega^+_\zeta$) and are called \textit{plus functions}. In addition to this any functions with a $-$ superscript are analytic in one of the regions, 
\begin{align}\label{QL-AK-minus-regions}
\Omega^-_z=\{z:|z|\geq|z_c|, z\neq z_c\},\quad \Omega^-_\zeta=\{\zeta:|\zeta|\geq|z_s|, \zeta\neq z_s\},
\end{align}
and are called \textit{minus functions}.
%\begin{remark}
%General Wiener--Hopf equations in two complex variables have recently been studied in diffraction theory for the quarter-plane problem \citep{AssierAbrahamsSIAP21} and the right-angled penetrable wedge problem \citep{KunzAssierSIAP2022}. In general, there are no known technique to solve them exactly. Nevertheless, information about the singularities of the unknown spectral functions can be unveiled through methods of analytic continuation \citep{AssierShaninQJMAM2018} \citep{KunzAssierQJMAM2023}, and recent development on asymptotic evaluation of double Fourier-type integrals \citep{AssierShaninKorolkovQJMAM2022} can be used to extract important information on the far-field behaviour of the physical field \citep{AssierShaninKorolkovQJMAM2024} \citep{KunzAssierIMA2024}. A similar approach can also be used for wave propagation within periodic structures \citep{ShaninAssierKorolkovMakarov2024}.
%
%The present Wiener--Hopf equation (\ref{QL-AK-WHE}) has a special form in the sense that some of its unknown functions depend on one variable only, allowing for mathematical progress as highlighted in \citep{Kisil2023}.
%\end{remark}

\subsection{Reduction to the Wiener--Hopf equation}

To proceed, we define a manifold such that $K(z,\zeta)=0$, which provides a relationship between the two variables. The function which describes this manifold, $\zeta=M(z)$, must satisfy $K(z,M(z))=0$ and is given by
\begin{align}\label{QL-AK-manifold-def}
M(z)=-\frac{L_1(z)}{2L_2(z)}+\sqrt{\frac{L_1(z)}{2L_2(z)}-1}\sqrt{\frac{L_1(z)}{2L_2(z)}+1},
\end{align}
where $L_1(z)$ and $L_2(z)$ are given in \eqref{QL-AK-kernelL1} and \eqref{QL-AK-kernelL2} respectively and the square roots take the principal branch\footnote{\label{QL-ft1}In general, we say that the principal branch of the square root $\sqrt{f(z)}$ (for any valid function $f(z)$) where $f(z)=|f|e^{i\varphi}$ restricts the argument of $f$, given by $\varphi$, to the range $(-\pi,\pi]$.}. Similarly $\zeta=1/M(z)$, $z=M(\zeta)$ and $z=1/M(\zeta)$ also satisfy $K(z,\zeta)=0$. Note that $L_1(z)$, $L_2(z)$ and $M(z)$ are reciprocally symmetric, i.e. satisfy the identity $M(z)=M(1/z)$ and $M(z)$ is its own inverse function (i.e. $M(M(z))=z$). In Appendix \ref{App:asympM}, we discuss the asymptotics of the singular points of $M(z)$, including a removable singularity and a pair of branch cuts. One of these branch cuts is inside the unit circle and has branch points at $z=M(1)$ and $z=M(-1)$. Due the reciprocal symmetry, the other cut is outside the unit circle and has branch points at $z=1/M(1)$ and $z=1/M(-1)$. If we put $\zeta=M(z)$ and separately $\zeta=1/M(z)$ in \eqref{QL-AK-WHE}, then we get two simplified Wiener--Hopf equations,
\begin{align}
\label{QL-AK-WHE-z+}-zL_2(z)B^{+}_1(z)-M(z)L_2(M(z))B^{+}_2(M(z))+A_{00}H^{(1)}_0\left(ks\sqrt{2}\right)&=zM(z)F^{++}(z,M(z)),\\
\label{QL-AK-WHE-z-}-zL_2(z)B^{+}_1(z)-\frac{1}{M(z)}L_2(M(z))B^{+}_2\left(\frac{1}{M(z)}\right)+A_{00}H^{(1)}_0\left(ks\sqrt{2}\right)&=\frac{z}{M(z)} F^{++}\left(z,\frac{1}{M(z)}\right).
\end{align}
We can then eliminate the unknowns $B^{+}_1(z)$ and $A_{00}$ by taking the difference between \eqref{QL-AK-WHE-z+} and \eqref{QL-AK-WHE-z-},
\begin{align}\nonumber\label{QL-AK-WHE-B2}
&M(z)L_2(M(z))B^{+}_2(M(z))+zM(z)F^{++}(z,M(z))\\
&=\frac{1}{M(z)}L_2(M(z))B^{+}_2\left(\frac{1}{M(z)}\right)+\frac{z}{M(z)} F^{++}\left(z,\frac{1}{M(z)}\right).
\end{align}
Next, we want to map \eqref{QL-AK-WHE-B2} back onto the $\zeta$ plane by putting $z=M(\zeta)$. Recalling that $M(\zeta)$ is its own inverse function, the Wiener--Hopf equation \eqref{QL-AK-WHE-B2} becomes,
%\begin{align}\label{QL-AK-WHE-B2-zeta}
%\zeta B^{+}_2(\zeta)-\frac{1}{\zeta}B^{+}_2\left(\frac{1}{\zeta}\right)=-\frac{M(\zeta)}{L_2(\zeta)}\left(\zeta \left(F^{++}(M(\zeta),\zeta)+ F_{\text{etc}}(M(\zeta),\zeta)\right)-\frac{1}{\zeta} \left(F^{++}\left(M(\zeta),\frac{1}{\zeta}\right)+ F_{\text{etc}}\left(M(\zeta),\frac{1}{\zeta}\right)\right)\right),
%\end{align}
\begin{align}\label{QL-WHE}
\zeta B^{+}_2(\zeta)-\frac{1}{\zeta}B^{+}_2\left(\frac{1}{\zeta}\right)
&=F_{\textrm{inc}}(\zeta)-F_{\textrm{inc}}\left(\frac{1}{\zeta}\right)+F_{A}(\zeta)-F_{A}\left(\frac{1}{\zeta}\right)
\end{align}
where
\begin{align}\label{QL-WHE-F}
F_{\textrm{inc}}(\zeta)&=\frac{M(\zeta)\zeta}{L_2(\zeta)(1-z_cM(\zeta))(1-z_s\zeta)},\quad%=-\zeta\frac{M(\zeta)}{L_2(\zeta)}F_{\text{etc}}(M(\zeta),\zeta)
F_{A}(\zeta)=\sum_{p,q=0}^{\infty}S_A(p,q)\frac{M(\zeta)^{p+1}\zeta^{q+1}}{L_2(\zeta)}.
\end{align}
The equation \eqref{QL-WHE} is a Wiener-Hopf equation in one complex variable. Indeed, $\zeta B^{+}_2(\zeta)$ is a plus function analytic in the region $\Omega^+_\zeta$, and $\displaystyle\frac{1}{\zeta}B^{+}_2\left(\frac{1}{\zeta}\right)$ is a minus function analytic in the region $\Omega^-_\zeta$.
All the forcing terms on the right-hand side are analytic in the intersection region $\Omega^+_\zeta\cap\Omega^-_\zeta$ that is an annulus that contains the unit circle contour $C_\zeta$. 

\subsection{Finding the Wiener--Hopf solution}\label{Sec:Solve}
To solve the Wiener--Hopf equation \eqref{QL-WHE} for the unknown function $B^{+}_2(\zeta)$, we just need to sum-split all the terms on the right-hand side because in this case, there is no kernel to factorise. Once solved, we then use \eqref{QL-AK-WHE-z+} to find $B^{+}_1(z)$ and then \eqref{QL-AK-WHE} to find $A^{++}(z,\zeta)$. Note that this will be an implicit solution as some unknown coefficients $A_{mn}$ are also present in $F_{A}(z,\zeta)$. 

\begin{figure}[ht]\centering
\includegraphics[width=0.5\textwidth]{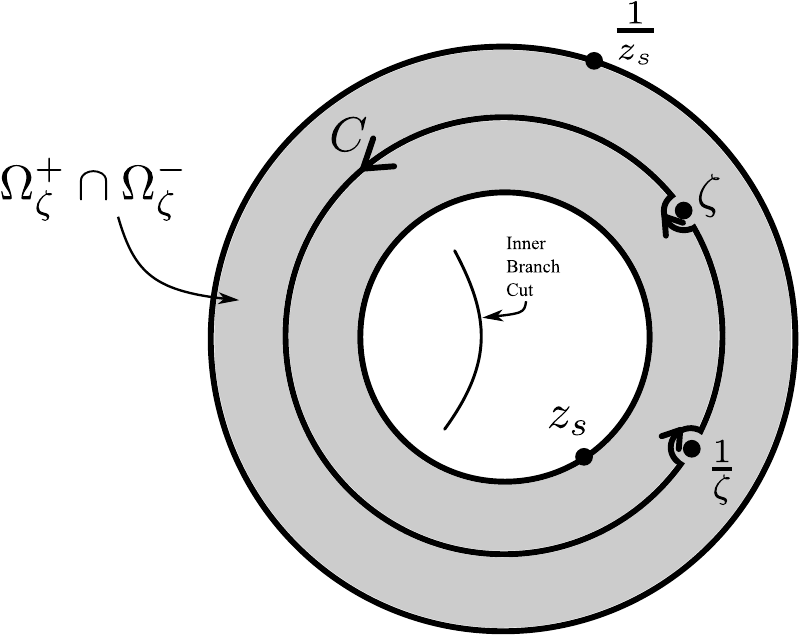}
\caption{Diagram of the Cauchy integral contour $C$ inside the grey intersection region $\Omega^+_\zeta\cap\Omega^-_\zeta$, while navigating around the poles at $z=\zeta$, $z=1/\zeta$, $z=z_s$ and $z=1/z_s$.}
\label{fig:QL-Cauchy-contour}
\end{figure}

For $F_{\textrm{inc}}(\zeta)$, we have a zero of order $2$ at $\zeta=0$, two simple poles at $\zeta=1/z_s$ and $\zeta=M(z_c)$ and two branch cuts coming from $M(\zeta)$. For $F_A(\zeta)$, each ($p,q$) term in the sum has a zero of order $q+2$ at $\zeta=0$, another two zeros of order $p$ (one inside and one outside the unit circle) where $\zeta L_2(\zeta)=0$ and the same two branch cuts from $M(\zeta)$. We sum-split the forcing terms $F_{\textrm{inc}}(\zeta)$ and $F_A(\zeta)$ given by \eqref{QL-WHE-F} using Cauchy's integral formula,
\begin{align}\nonumber
F(\zeta)&=F^+(\zeta)+F^-(\zeta),\\
&=\frac{1}{2\pi i}\int_{C}\frac{F\left(\frac{1}{z}\right)}{z(1-\zeta z)}\text{d}z-\frac{1}{2\pi i}\int_{C}\frac{F(z)}{z-\zeta}\text{d}z,
\end{align}
where the contour $C$ is an anticlockwise contour on the unit circle contained in the annulus $\Omega^+_\zeta\cap\Omega^-_\zeta$, which passes radially below poles at $z=\zeta$ and $z=1/\zeta$ as well as a possible pole at $z=1/z_s$ while also passing radially above another possible pole at $z=z_s$ (see Figure \ref{fig:QL-Cauchy-contour} for diagram).

We can then rewrite the Wiener--Hopf equation \eqref{QL-WHE} as,
\begin{align}\nonumber\label{QL-WHE-final}
\zeta B^{+}_2(\zeta)-&\left[F_{\text{inc}}^+(\zeta)-F_{\text{inc}}^-\left(\frac{1}{\zeta}\right)+F_{A}^+(\zeta)-F_{A}^-\left(\frac{1}{\zeta}\right)\right]\\
=&\frac{1}{\zeta}B^{+}_2\left(\frac{1}{\zeta}\right)+\left[F_{\text{inc}}^-(\zeta)-F_{\text{inc}}^+\left(\frac{1}{\zeta}\right)+F_{A}^-(\zeta)-F_{A}^+\left(\frac{1}{\zeta}\right)\right],
\end{align}
where all the terms on the left-hand (resp. right-hand) side are analytic in the region $\Omega^+_\zeta$ (resp. $\Omega^-_\zeta$) and the Wiener--Hopf equation is valid on the annulus $\Omega^+_\zeta\cap\Omega^-_\zeta$. We can create an entire function, which we call $\Psi(\zeta)$, defined as follows
\begin{align}\label{QL-WHE-Psi}
\Psi(\zeta)=\begin{cases}
\textrm{LHS} \eqref{QL-WHE-final}& \textrm{inside},\\
\textrm{RHS} \eqref{QL-WHE-final}& \textrm{outside},\\
\textrm{LHS} \eqref{QL-WHE-final} = \textrm{RHS} \eqref{QL-WHE-final}& \textrm{annulus}.
\end{cases}
%O\left(\frac{1}{\zeta}\right)+O\left(\frac{1}{\zeta}\right)-O(1)+O\left(\frac{1}{\zeta}\right)-O(1)
\end{align}
Among the terms on the right-hand side of \eqref{QL-WHE-final}, we find that $F_{\text{inc}}^+\left(\frac{1}{\zeta}\right)$ and $F_{A}^+\left(\frac{1}{\zeta}\right)$ are $O(1)$ as $|\zeta|\rightarrow\infty$ and the others are $O\left(\frac{1}{\zeta}\right)$. This means that the entire function is $O(1)$ for large $|\zeta|$, so by applying Liouville's theorem, we conclude that $\Psi(\zeta)$ is a constant given by,
\begin{align}\label{QL-WHE-Psi-constant}
\lim_{|\zeta|\rightarrow\infty}\Psi(\zeta)=-F_{\text{inc}}^+(0)-F_{A}^+(0)=-\frac{1}{2\pi i}\int_{C}\frac{F_{\text{inc}}^+\left(\frac{1}{z}\right)+F_{A}^+\left(\frac{1}{z}\right)}{z}\text{d}z.
\end{align}
This means that the solution to the WH equation \eqref{QL-WHE-final} is
\begin{align}\label{QL-B2-sol}
B^{+}_2(\zeta)=\frac{1}{\zeta}\left[F_{\text{inc}}^+(\zeta)-F_{\text{inc}}^-\left(\frac{1}{\zeta}\right)-F_{\text{inc}}^+(0)+F_{A}^+(\zeta)-F_{A}^-\left(\frac{1}{\zeta}\right)-F_{A}^+(0)\right].
\end{align}
We can combine the different Cauchy integrals so that equation \eqref{QL-B2-sol} can be decomposed into two parts of the form
\begin{align}
\frac{1}{\zeta}\left[F^+(\zeta)-F^-\left(\frac{1}{\zeta}\right)-F^+(0)\right]&=
\frac{1}{2\pi i}\int_{C}\frac{F\left(\frac{1}{z}\right)-F(z)}{1-\zeta z}\text{d}z,
%&=\frac{1}{2\pi}\int_{-\pi}^{\pi}\frac{F\left(e^{it}\right)}{e^{it}-\zeta}\text{d}t+\frac{1}{2\pi}\int_{-\pi}^\pi\frac{F(e^{it})e^{it}}{\zeta e^{it}-1}\text{d}t,
\end{align}
where we can substitute $F=F_{\text{inc}}$ or $F_{A}$. This means that the explicit solution for $B_2^+(\zeta)$ is given by,
\begin{align}\nonumber\label{QL-B2-sol-integral}
B_2^+(\zeta)=\ &\frac{1}{2\pi i}\int_{C}\frac{M(z)}{L_2(z)(1-z_cM(z))(1-\zeta z)}\left(\frac{1}{z-z_s}-\frac{z}{1-z_sz}\right)\text{d}z\\
&+\sum_{p,q=0}^{\infty}\frac{S_A(p,q)}{2\pi i}\int_{C}\frac{M(z)^{p+1}}{L_2(z)(1-\zeta z)}\left(z^{-q-1}-z^{q+1}\right)\text{d}z.
\end{align}

%\begin{align}\nonumber\label{QL-B2-sol-numerical}
%B_2^+(\zeta)=\ &\frac{1}{2\pi}\int_{C}\frac{\mathcal{F}(z)}{1-\zeta z}\left(\frac{1}{z-z_s}-\frac{z}{1-z_sz}\right)
%-\frac{\mathcal{F}(\zeta)}{1-\zeta z}\left(\frac{\zeta}{1-z_s\zeta}-\frac{1}{\zeta-z_s}\right)-\frac{\mathcal{F}(z_s)}{(1-z_s\zeta)(z-z_s)}\\
%\nonumber&-\frac{\mathcal{F}(z_s)}{(\zeta-z_s)(1-z_sz)}\text{d}z+\frac{\mathcal{F}(z_s)}{1-z_s\zeta}
%+\sum_{p,q=0}^{\infty}S_A(p,q)\bigg[\\
%&\frac{1}{2\pi}\int_{C}\frac{1}{1-\zeta z}\left(\frac{M(z)^{p+1}}{L_2(z)}\left(z^{-q-1}-z^{q+1}\right)
%+\frac{M(\zeta)^{p+1}}{L_2(\zeta)}\left(\zeta^{-q-1}-\zeta^{q+1}\right)\right)\text{d}z\bigg],
%\end{align}
%where
%\begin{align}
%\mathcal{F}(z)=\frac{M(z)}{L_2(z)(1-z_cM(z))}.
%\end{align}

We can write a formula for $B_1^{+}(z)$ in terms of $B_2^+$ by rearranging \eqref{QL-AK-WHE-z+},
\begin{align}\label{QL-B1-sol}
zL_2(z)B^{+}_1(z)-A_{00}H^{(1)}_0\left(ks\sqrt{2}\right)=-M(z)L_2(M(z))B^{+}_2(M(z))-zM(z)F^{++}(z,M(z)),
\end{align}
which we then substitute into \eqref{QL-AK-WHE} to write $A^{++}(z,\zeta)$ in terms of $B_2^+(\zeta)$, 
\begin{align}\label{QL-A++-sol}
A^{++}(z,\zeta)%=&\frac{1}{K(z,\zeta)}\bigg[\frac{1}{\zeta}\left(L_2(z)B^{+}_1(z)-\frac{A_{00}}{z}\right)+\frac{1}{z}L_2(\zeta)B^{+}_2(\zeta) +F^{++}(z,\zeta)\bigg],\\
=&\frac{1}{K(z,\zeta)}\bigg[\frac{1}{z}L_2(\zeta)B^{+}_2(\zeta)-\frac{M(z)}{z\zeta}L_2(M(z))B^{+}_2(M(z))
-\frac{M(z)}{\zeta}F^{++}(z,M(z))+F^{++}(z,\zeta)\bigg],
\end{align}

To evaluate \eqref{QL-B2-sol-integral} numerically, we must take care to avoid accidentally crossing the branch cuts of $M(z)$. However, the associated branch points do not cross the unit circle and, in most cases, they are not even in the neighbourhood of $C$ so it is very unlikely that deformations are required. There are also two simple poles where $L_2(z)=0$, but these are removable singularities that are cancelled out by the removable singularities of $M(z)$ (see Appendix \ref{App:asympM}) in the numerator which are at the same locations and behave like simple zeros.

For the $F_{\text{inc}}$ part, there are simple poles at $z=M\left(1/z_c\right)$ and $z=1/M\left(1/z_c\right)$, but these are located on the branch cuts and do not need to be avoided. There are also three simple poles at $z=z_s$, $z=1/z_s$ and $z=1/\zeta$ in the integrand that \textit{can} interact with the integration contour $C$. The first two are positioned near the integration contour and the last can cross it. Of all these poles, only the one at $z=z_s$ is inside the contour. 

For the $F_{A}$ part, there is a pole of order $q+1$ at $z=0$ and a simple pole at $z=1/\zeta$, but only the latter can cross the integration contour. To numerically evaluate the Cauchy integrals efficiently, we use the pole removal technique\footnote{\label{QL-ft2}The pole removal technique consists in adding and subtracting a pole inside the integrand to split the integral into two parts; one non-singular integral simple to evaluate numerically and an integral of the removed pole which can be evaluated exactly.} on all of the simple poles that can interact with the integration contour. This leads to integrands with no singularities on or near the integration contour. In the end, $B_2^+(\zeta)$ will have one simple pole at $\zeta=1/z_s$ and no singularities within the region $\Omega^+_\zeta$.

\subsection{Finding the scattering coefficients}\label{Sec:Amn}
To find the scattering coefficients we let the imaginary part of $k$ tend to zero and apply the inverse transform given by \eqref{QL-Z-trans-inv}, onto the Wiener--Hopf solution \eqref{QL-A++-sol}. We tackle each double integral separately by decomposing \eqref{QL-Z-trans-inv} into four parts, $A_{mn}=A_{mn}^{(1)}+A_{mn}^{(2)}+A_{mn}^{(3)}+A_{mn}^{(4)}$, where,
\begin{align}
\label{QL-Z-trans-inv-1}A_{mn}^{(1)}&=
\frac{1}{(2\pi i)^2}\int_{C_z}\int_{C_\zeta}\frac{z^{-m-2}\zeta^{-n-1}}{K(z,\zeta)}L_2(\zeta)B^{+}_2(\zeta)\textrm{d}\zeta\textrm{d}z,\\
\label{QL-Z-trans-inv-2}A_{mn}^{(2)}&=
-\frac{1}{(2\pi i)^2}\int_{C_z}\int_{C_\zeta}\frac{M(z)z^{-m-2}\zeta^{-n-2}}{K(z,\zeta)}L_2(M(z))B^{+}_2(M(z))\textrm{d}\zeta\textrm{d}z,\\
\label{QL-Z-trans-inv-3}A_{mn}^{(3)}&=
-\frac{1}{(2\pi i)^2}\int_{C_z}\int_{C_\zeta}\frac{M(z)z^{-m-1}\zeta^{-n-2}}{K(z,\zeta)}F^{++}(z,M(z))\textrm{d}\zeta\textrm{d}z,\\
\label{QL-Z-trans-inv-4}A_{mn}^{(4)}&=
\frac{1}{(2\pi i)^2}\int_{C_z}\int_{C_\zeta}\frac{z^{-m-1}\zeta^{-n-1}}{K(z,\zeta)}F^{++}(z,\zeta)\textrm{d}\zeta\textrm{d}z,
\end{align}

Before evaluating the above, we will aim to simplify these integrals as much as possible. One reason for this is that these integrals contain unknown $A_{mn}$ within them and by simplifying, we are able to reduce it to a linear system \eqref{QL-A-sol}. The second reason is that the above is computationally costly to compute given that the expression for \(B^{+}_2\) involves sums of integrals. 

In each of these double integrals, one of the integrals can be evaluated analytically. For \eqref{QL-Z-trans-inv-1}, we evaluate the $z$ integral first and for \eqref{QL-Z-trans-inv-2} and \eqref{QL-Z-trans-inv-3}, we evaluate the $\zeta$ integral first using residue theorem (see Appendix~\ref{App:integral})
\begin{align}\label{QL-A++-int123}
\frac{1}{2\pi i}\int_{C_\zeta}\frac{\zeta^{-n-2}}{K(z,\zeta)}\textrm{d}\zeta&=\frac{M(z)^{n+2}}{L_2(z)(M(z)^2-1)}.
\end{align}

For \eqref{QL-Z-trans-inv-4}, we have the option to chose which integral we first evaluate analytically and we picked the $\zeta$ integral. The double integral \eqref{QL-Z-trans-inv-4} is also in two parts. For one part we use the following formula
\begin{align}\label{QL-A++-int4-part1}
\frac{1}{2\pi i}\int_{C_\zeta}\frac{\zeta^{-n-1}}{K(z,\zeta)(z_s\zeta-1)}\textrm{d}\zeta
&=-\frac{M(z)}{L_2(z)(M(z)-z_s)}\left[\frac{M(z)^{n+1}}{M(z)^2-1}-\frac{z_s^{n+1}}{(z_sM(z)-1)}\right].
\end{align}
For the other part, the $\zeta$ integral simplifies to something similar to \eqref{QL-A++-int123}
\begin{align}\label{QL-A++-int4-part2}
\frac{1}{2\pi i}\int_{C_\zeta}\frac{\zeta^{q-n-1}}{K(z,\zeta)}\textrm{d}\zeta
%\nonumber&=\frac{1}{2\pi i}\int_{C_\zeta}\frac{\zeta^{q-n}}{L_2(z)(\zeta-M(z))(\zeta-M(z)^{-1})}\textrm{d}\zeta\\
%\nonumber&=\underset{\zeta=M(z)}{\text{Res}}\left(\frac{\zeta^{q-n-1}}{K(z,\zeta)}\right)
%+\mathcal{H}(n-q)\underset{\zeta=0}{\text{Res}}\left(\frac{\zeta^{q-n-1}}{K(z,\zeta)}\right)\\
%\nonumber&=\frac{M(z)^{q-n+1}}{L_2(z)(M(z)^2-1)}+\mathcal{H}(n-q)\frac{M(z)^{n-q+1}-M(z)^{q-n+1}}{L_2(z)(M(z)^2-1)}\\
&=\frac{M(z)^{|n-q|+1}}{L_2(z)(M(z)^2-1)},
\end{align}
%where $\mathcal{H}(z)$ is the Heaviside function. 
Again, details on the evaluation of the three integrals \eqref{QL-A++-int123}, \eqref{QL-A++-int4-part1} and \eqref{QL-A++-int4-part2} is given in Appendix \ref{App:integral}.
%After we apply these formulae to $A_{mn}$,
%\begin{align}
%\label{QL-Amn-part1}A_{mn}=&\frac{1}{2\pi i}\int_{C_\zeta}\frac{M(\zeta)^{m+2}\zeta^{-n-1}}{M(\zeta)^2-1}B^{+}_2(\zeta)\textrm{d}\zeta\\
%\label{QL-Amn-part2}&-\frac{1}{2\pi i}\int_{C_z}\frac{L_2(M(z))M(z)^{n+3}z^{-m-2}}{L_2(z)(M(z)^2-1)}B^{+}_2(M(z))\textrm{d}z\\
%\label{QL-Amn-part3}&+\frac{1}{2\pi i}\int_{C_z}\frac{M(z)^{n+3}z^{-m-1}}{L_2(z)(M(z)^2-1)(1-z_cz)(1-z_sM(z))}\textrm{d}z\\
%\label{QL-Amn-part4}&+\sum_{p,q=0}^{\infty}S_A(p,q)\frac{1}{2\pi i}\int_{C_z}\frac{M(z)^{n+q+3}z^{p-m-1}}{L_2(z)(M(z)^2-1)}\textrm{d}z\\
%\label{QL-Amn-part5}&+\frac{1}{2\pi i}\int_{C_z}\frac{M(z)z^{-m-1}}{L_2(z)(M(z)-z_s)(1-z_cz)} \left[\frac{z_s^{n+1}}{(z_sM(z)-1)}-\frac{M(z)^{n+1}}{M(z)^2-1}\right]\textrm{d}z\\
%\label{QL-Amn-part6}&-\sum_{p,q=0}^{\infty}S_A(p,q)\frac{1}{2\pi i}\int_{C_z}\frac{M(z)^{|n-q|+1}z^{p-m-1}}{L_2(z)(M(z)^2-1)}\textrm{d}z.
%\end{align}
%where, 
%\begin{align}\label{QL-AK-SumA}
%S_A(p,q)=\sum_{(\bar{m},\bar{n})\in \bar{S}_{pq}}A_{\bar{m}\bar{n}}H^{(1)}_0(ks\sqrt{(p-\bar{m})^2+(q-\bar{n})^2}).
%\end{align} 
%Here, we can combine \eqref{QL-Amn-part3} and \eqref{QL-Amn-part5}, and also combine \eqref{QL-Amn-part4} and \eqref{QL-Amn-part6}.

Once \eqref{QL-A++-int123} has been used, the integral\eqref{QL-Z-trans-inv-2} reduces to,
\begin{align}
A_{mn}^{(2)}=-\frac{1}{2\pi i}\int_{C_z}\frac{L_2(M(z))M(z)^{n+3}z^{-m-2}}{L_2(z)(M(z)^2-1)}B^{+}_2(M(z))\textrm{d}z.
\end{align}
Using the substitution $z=1/M(\zeta)$, it becomes
\begin{align}
A_{mn}^{(2)}=&\frac{1}{2\pi i}\int_{\Gamma_\zeta}\frac{M'(\zeta)L_2(\zeta)M(\zeta)^{m}\zeta^{n+3}}{L_2(M(\zeta))(\zeta^2-1)}B^{+}_2(\zeta)\textrm{d}\zeta,
\end{align}
which can be simplified by noting that,
\begin{align}\label{QL-AK-manifold-der}
M'(\zeta)=-\frac{M(\zeta)^2L_2(M(\zeta))(\zeta^2-1)}{\zeta^2L_2(\zeta)(M(\zeta)^2-1)},
\end{align}
leading to,
\begin{align}\label{QL-Z-trans-inv-2-Gamma}
A_{mn}^{(2)}=&-\frac{1}{2\pi i}\int_{\Gamma_\zeta}\frac{M(\zeta)^{m+2}\zeta^{n+1}}{M(\zeta)^2-1}B^{+}_2(\zeta)\textrm{d}\zeta.
\end{align}
The contour $\Gamma_\zeta$ is the image of the contour $C_z$ by the $z=1/M(\zeta)$ substitution and it encircles the branch cut inside the unit circle in an anticlockwise direction. %It also demonstrates how a possible pole that was originally on the unit circle is affected by this transform, however this does not apply to $A_{mn}^{(2)}$.
\begin{figure}[ht]\centering
\includegraphics[width=0.75\textwidth]{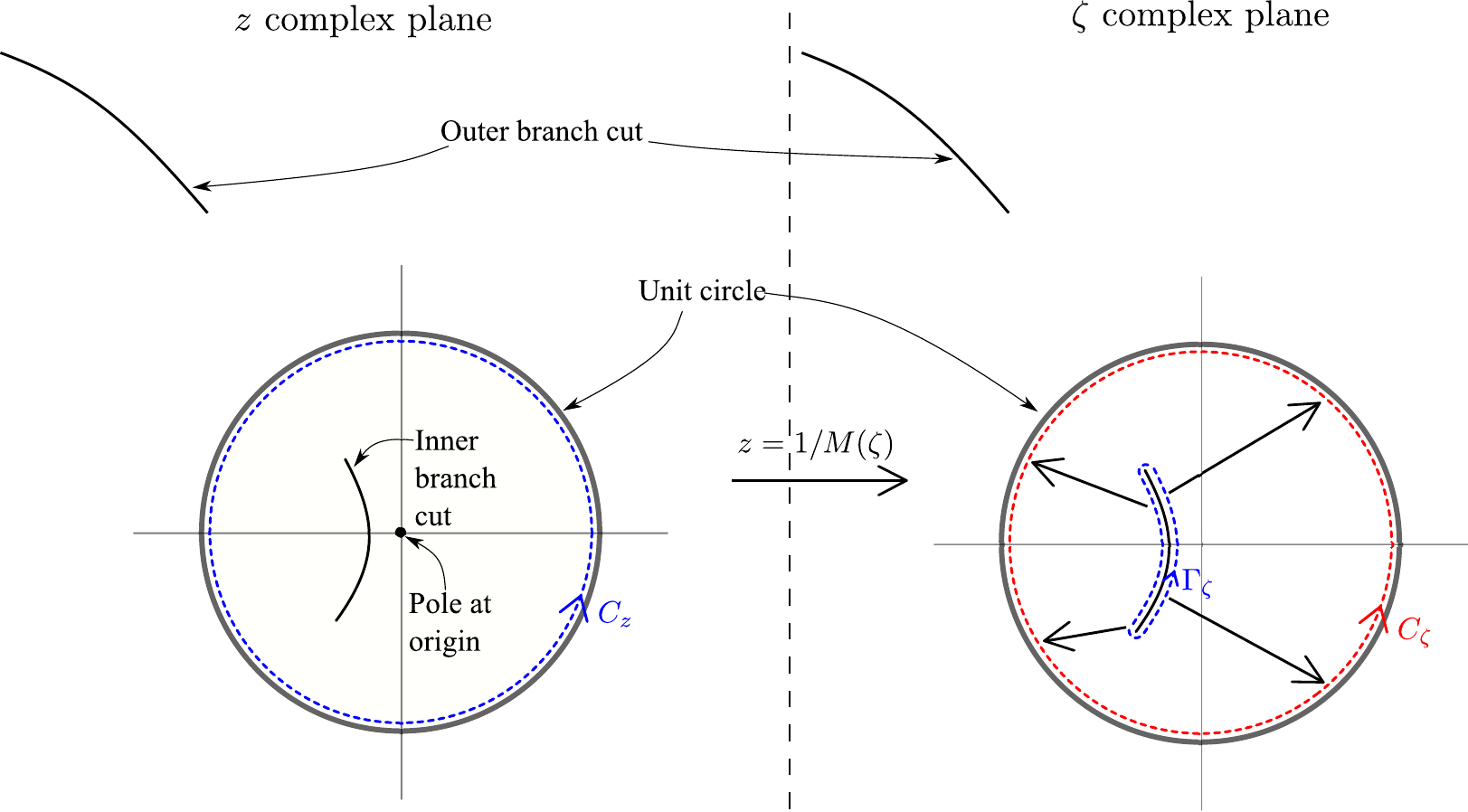}
\caption{This is a diagram of the integration contours $C_z$, $C_{(1/M(\zeta))}$ and $C_\zeta$.}
\label{fig:QL-int-sub-contours}
\end{figure}
The new integrand of \eqref{QL-Z-trans-inv-2-Gamma} does not have any singularities between $\Gamma_\zeta$ and $C_\zeta$ (although $B^{+}_2(\zeta)$ does have an simple pole at $\zeta=1/z_s$ that is outside $C_\zeta$), so we can freely deform the contour $\Gamma_\zeta$ to the unit circle $C_\zeta$ without crossing any singularities (by making the deformed contour pass radially below the pole at $\zeta=1/z_s$), leading to
\begin{align}
A_{mn}^{(2)}=&-\frac{1}{2\pi i}\int_{C_{\zeta}}\frac{M(\zeta)^{m+2}\zeta^{n+1}}{M(\zeta)^2-1}B^{+}_2(\zeta)\textrm{d}\zeta.
\end{align}
Figure \ref{fig:QL-int-sub-contours} illustrates the transformation of the integration contours from $C_z$ to $\Gamma_{\zeta}$ via the substitution $z=1/M(\zeta)$ before we deform $\Gamma_\zeta$ to $C_\zeta$.

After all these integral evaluations and substitutions, we can combine the single remaining integrals of $A_{mn}^{(1)}$ and $A_{mn}^{(2)}$ as well as combine the remaining integrals of $A_{mn}^{(3)}$ and $A_{mn}^{(4)}$ to obtain,
\begin{align}\nonumber\label{QL-Amn-allparts}
A_{mn}=&\frac{1}{2\pi i}\int_{C_\zeta}\frac{M(\zeta)^{m+2}}{M(\zeta)^2-1}\left(\zeta^{-n-1}-\zeta^{n+1}\right)B^{+}_2(\zeta)\textrm{d}\zeta\\
\nonumber&+\frac{1}{2\pi i}\int_{C_z}\frac{M(z)z^{-m-1}(M(z)^{n+1}-z_s^{n+1})}{L_2(z)(M(z)-z_s)(1-z_sM(z))(1-z_cz)}\textrm{d}z\\
&+\sum_{p,q=0}^{\infty}\frac{S_A(p,q)}{2\pi i}\int_{C_z}\frac{z^{p-m-1}\left(M(z)^{n+q+3}-M(z)^{|n-q|+1}\right)}{L_2(z)(M(z)^2-1)}\textrm{d}z.
\end{align}
For the first integral in \eqref{QL-Amn-allparts}, the only singularity of concern is the one at $\zeta=1/z_s$ from $B^{+}_2(\zeta)$ which is outside the contour so the residue is not included. Hence, we can numerically evaluate this integral efficiently using the pole removal technique$^{\ref{QL-ft2}}$. For the second integral of \eqref{QL-Amn-allparts}, we also use the pole removal technique on the pole at $z=1/z_c$ which is also outside the contour so the residue is not included.
%For the second integral in \eqref{QL-Amn-allparts}, it is possible to use the substitution $z=1/M(\zeta)$ to simplify the integrand, however there is a pole $z=1/z_c$ that will be crossed when deforming the transformed contour to the unit circle $C_\zeta$ (also note that we will need to deal with another pole at $\zeta=1/z_s$).
The last integral of \eqref{QL-Amn-allparts} can be shown to be symmetric about $n=q$. We could also use the substitutions $z=M(\zeta)$ or $z=1/M(\zeta)$ to simplify the integrand, the results of which show that this integral is symmetric about $m=p$. Hence, we can rewrite the last integral in \eqref{QL-Amn-allparts} as follows,
\begin{align}\label{QL-Amn-lastpart}
\int_{C_\zeta}\frac{M(\zeta)^{|m-p|+1}\left(\zeta^{n+q+1}-\zeta^{|n-q|-1}\right)}{L_2(\zeta)(M(\zeta)^2-1)}\textrm{d}\zeta,
\end{align}
where we used \eqref{QL-AK-manifold-der} to simplify the transformed integrand and remove another occurrence of $M'(\zeta)$.

\begin{figure}[ht]\centering
\includegraphics[width=0.7\textwidth]{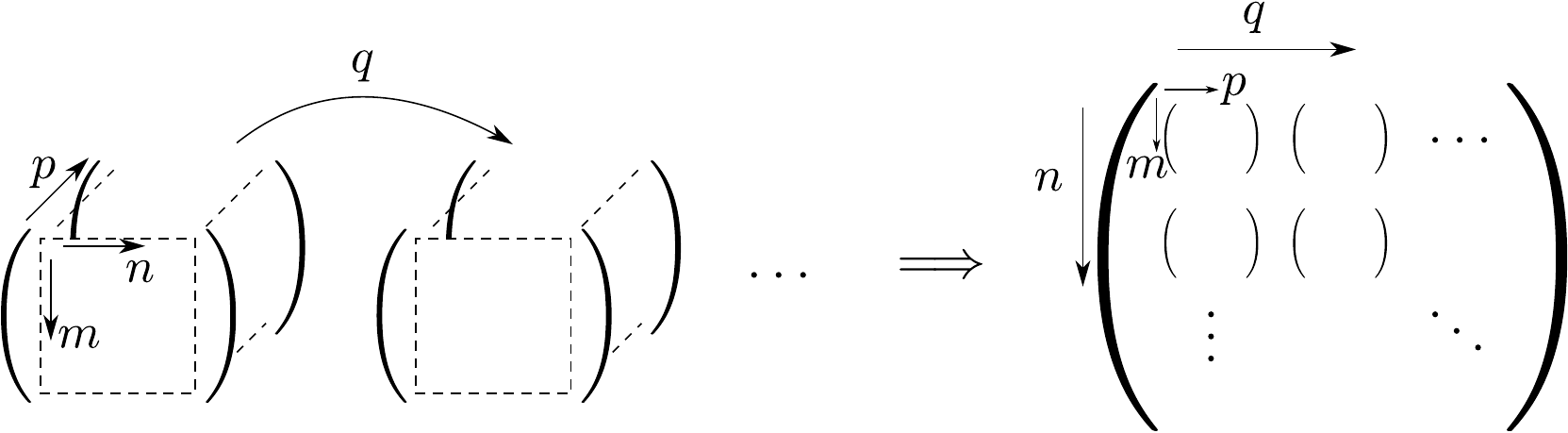}
\caption{Diagram of how we reshape a four-dimensional tensor with indices $m,n,p,q$ to a two-dimensional $n\times q$ block matrix with blocks of size $m\times p$. Note that $A_{mn}$ is reshaped in the same way from a two-dimensional matrix to a column vector.}
\label{fig:QL-reshape_4d_to_2d}
\end{figure}

In the end, we can substitute the formula for $B_2^+(\zeta)$ given by \eqref{QL-B2-sol-integral} and then $A_{mn}$ is given by,
\begin{align}\nonumber\label{QL-Amn-inc+etc-parts}
A_{mn}=&\ \frac{1}{(2\pi i)^2}\!\int_{C_\zeta}\!\!\frac{M(\zeta)^{m+2}}{M(\zeta)^2-1}\!\left(\zeta^{-n-1}\!-\!\zeta^{n+1}\right)\!\!
\int_{C_z}\!\!\frac{M(z)}{L_2(z)(1-z_cM(z))(1-\zeta z)}\!\left(\!\frac{1}{(z-z_s)}\!-\!\frac{z}{(1-z_sz)}\!\right)\!\text{d}z\textrm{d}\zeta\\
\nonumber&+\frac{1}{2\pi i}\int_{C_z}\frac{M(z)z^{-m-1}(M(z)^{n+1}-z_s^{n+1})}{L_2(z)(M(z)-z_s)(1-z_sM(z))(1-z_cz)}\textrm{d}z\\
\nonumber&+\sum_{p,q=0}^{\infty}\frac{S_A(p,q)}{2\pi i}\Bigg[\int_{C_\zeta}\frac{M(\zeta)^{m+2}}{M(\zeta)^2-1}\left(\zeta^{-n-1}-\zeta^{n+1}\right)
\frac{1}{2\pi i}\int_{C_z}\frac{M(z)^{p+1}}{L_2(z)(1-\zeta z)}\left(z^{-q-1}-z^{q+1}\right)\text{d}z\textrm{d}\zeta\\
&+\int_{C_\zeta}\frac{M(\zeta)^{|m-p|+1}\left(\zeta^{n+q+1}-\zeta^{|n-q|-1}\right)}{L_2(\zeta)(M(\zeta)^2-1)}\textrm{d}\zeta\Bigg],
\end{align}
where $S_A(p,q)$ is defined in \eqref{QL-AK-WHE-forcing}. 
%$$S_A(p,q)=\sum_{(\bar{m},\bar{n})\in \bar{S}_{pq}}A_{\bar{m}\bar{n}}H^{(1)}_0(ks\sqrt{(p-\bar{m})^2+(q-\bar{n})^2}).$$
In this format, the first two lines of \eqref{QL-Amn-inc+etc-parts} contain the influence of the incident wave on the scattering coefficients, whereas the last two lines contain the influence of the scattering coefficients outside the initial $3\times3$ block of scatterers shown in Figure \ref{fig:QL-highlighted}. This format also shows that by the following rearrangement of sums,
$$\sum_{p,q=0}^{\infty}\sum_{(\bar{m},\bar{n})\in \bar{S}_{pq}}\equiv\sum_{\bar{m},\bar{n}=0}^{\infty}\sum_{(p,q)\in \bar{S}_{\bar{m}\bar{n}}},$$
then we can write \eqref{QL-Amn-inc+etc-parts} using suffix notation,
\begin{align}\label{QL-tensor-form}
A_{mn}=A^{(I)}_{mn}+\mathcal{M}^{(1)}_{mnpq}\mathcal{M}^{(2)}_{pq\bar{m}\bar{n}}A_{\bar{m}\bar{n}}.
\end{align}
Here, the values of $A^{(I)}_{mn}$ are given by,
\begin{align}\nonumber\label{QL-A-inc-part}
A^{(I)}_{mn}=&\ \frac{1}{(2\pi i)^2}\!\int_{C_\zeta}\!\!\frac{M(\zeta)^{m+2}}{M(\zeta)^2-1}\!\left(\zeta^{-n-1}\!-\!\zeta^{n+1}\right)\!\!
\int_{C_z}\!\!\frac{M(z)}{L_2(z)(1-z_cM(z))(1-\zeta z)}\!\left(\!\frac{1}{(z-z_s)}\!-\!\frac{z}{(1-z_sz)}\!\right)\!\text{d}z\textrm{d}\zeta\\
&+\frac{1}{2\pi i}\int_{C_z}\frac{M(z)z^{-m-1}(M(z)^{n+1}-z_s^{n+1})}{L_2(z)(M(z)-z_s)(1-z_sM(z))(1-z_cz)}\textrm{d}z,
\end{align}
the values of $\mathcal{M}^{(1)}_{mnpq}$ are given by,
\begin{align}\nonumber\label{QL-A-M1}
\mathcal{M}^{(1)}_{mnpq}=\frac{1}{2\pi i}\Bigg[&\int_{C_\zeta}\frac{M(\zeta)^{m+2}}{M(\zeta)^2-1}\left(\zeta^{-n-1}-\zeta^{n+1}\right)
\frac{1}{2\pi i}\int_{C_z}\frac{M(z)^{p+1}}{L_2(z)(1-\zeta z)}\left(z^{-q-1}-z^{q+1}\right)\text{d}z\textrm{d}\zeta\\
&+\int_{C_\zeta}\frac{M(\zeta)^{|m-p|+1}\left(\zeta^{n+q+1}-\zeta^{|n-q|-1}\right)}{L_2(\zeta)(M(\zeta)^2-1)}\textrm{d}\zeta\Bigg],
\end{align}
and the values of $\mathcal{M}^{(2)}_{pq\bar{m}\bar{n}}$ are given by,
\begin{align}\label{QL-A-M2}
\mathcal{M}^{(2)}_{pq\bar{m}\bar{n}}=\begin{cases}
%H^{(1)}_0(ks\sqrt{(p-\bar{m})^2+(q-\bar{n})^2})
0 & (p,q) \in S_{\bar{m}\bar{n}},\\
\mathbb{H}_{(p-\bar{m})(q-\bar{n})}& (p,q) \notin S_{\bar{m}\bar{n}}.
\end{cases}
\end{align}
If we reshape the scattering coefficient matrix to a column vector and reshape the 4D tensors to 2D matrices (see Figure \ref{fig:QL-reshape_4d_to_2d} for diagram),
then we can produce a standard infinite matrix equation which can be inverted. The solution to this infinite matrix equation is given by
\begin{align}\label{QL-A-sol}
%A=A^{(I)}+\mathcal{M}A, \implies 
A=(\mathcal{I}-\mathcal{M})^{-1}A^{(I)},
\end{align}
where $A$ (resp. $A^{\RED{(I)}}$) is $A_{mn}$ (resp. $A^{\RED{(I)}}_{mn}$) written as a column vector, $\mathcal{M}$ is $\mathcal{M}^{(1)}_{mnpq}\mathcal{M}^{(2)}_{pq\bar{m}\bar{n}}$ written as a block matrix and $\mathcal{I}$ is the identity matrix.

Note that when we numerically evaluate the inner integrals of $\mathcal{M}^{(1)}_{mnpq}$, we remove the pole at $z=1/\zeta$ in the integrand but do not add the residue because the pole is outside the closed contour $C_z$. However, when $p,q\rightarrow\infty$ or $m,n\rightarrow\infty$, we can also use the following behaviours for the inner integrals
\begin{align}
\label{QL-A-M1-asymp_pq}\frac{1}{2\pi i}\int_{C_z}\frac{M(z)^{p+1}}{L_2(z)(1-\zeta z)}(z^{-q-1}-z^{q+1})\textrm{d}z&\underset{p,q\rightarrow\infty}{\sim}\frac{M(\zeta)^{p+1}\zeta^q}{L_2(\zeta)},\\
\label{QL-A-M1-asymp_mn}\frac{1}{2\pi i}\int_{C_\zeta}\frac{M(\zeta)^{m+2}}{(M(\zeta)^2-1)(1-\zeta z)}(\zeta^{-n-1}-\zeta^{n+1})\textrm{d}\zeta &\underset{m,n\rightarrow\infty}{\sim}\frac{M(z)^{m+2}z^n}{(M(z)^2-1)},
\end{align}
provided that $|\zeta|\geq1$, which comes from the residue at $z=1/\zeta$ when you let the radius of $C_z$ grow to infinity. We use this behaviour to show that the asymptotic behaviour of the values in $\mathcal{M}^{(1)}_{mnpq}$ as $m,n,p,q\rightarrow\infty$ is the single integral,
\begin{align}\label{QL-A-M1-asymp}
\mathcal{M}^{(1)}_{mnpq}\sim\ &\frac{1}{2\pi i}\int_{C_\zeta}\frac{M(\zeta)^{m+p+3}-M(\zeta)^{|m-p|+1}}{L_2(\zeta)(M(\zeta)^2-1)}(\zeta^{|n-q|-1}-\zeta^{n+q+1})\textrm{d}\zeta,
%\sim&\frac{1}{2\pi i}\int_{C_z}\frac{M(z)^{m+p+3}}{L_2(z)(M(z)^2-1)}(z^{n-q-1}-z^{n+q+1})\textrm{d}z\\
%&+\frac{1}{2\pi i}\int_{C_\zeta}\frac{M(\zeta)^{|m-p|+1}\left(\zeta^{n+q+1}-\zeta^{|n-q|-1}\right)}{L_2(\zeta)(M(\zeta)^2-1)}\textrm{d}\zeta.
\end{align}
and doing so provides a faster calculation for the values of $\mathcal{M}^{(1)}_{mnpq}$ while still being accurate. On the subject of saving computation time, we can also create rational approximations of the inner integrals in $\mathcal{M}^{(1)}_{mnpq}$ and $A^{\RED{(I)}}_{mn}$ on the contour and use the result for a highly accurate and significantly quicker-to-compute approximation. We create these rational approximations in MATLAB using the ``AAA'' function in Chebfun \citep{Nakatsukasa2018}. While for $A^{\RED{(I)}}_{mn}$, we only require one rational approximation, for $\mathcal{M}^{(1)}_{mnpq}$, we require an approximation for every pair of $(p,q)$. So in our numerical calculations, we use rational approximations when $m,n,p$ and $q$ are small, but if one of them is above a chosen threshold, we use \eqref{QL-A-M1-asymp} to calculate $\mathcal{M}^{(1)}_{mnpq}$ instead.

\section{Comparisons and results}\label{Sec:Results}
In this section, we consider several test cases, for which we plot the real part of the total field. Our results, labelled QLNN (standing for `Quarter Lattice Nearest Neighbours'), are critically compared with three other methods: MSIA (standing for `Multiple Semi-Infinite Arrays') referring to the method and solution described in our previous article \citep{MSIApaper}; TMAT referring to the T-matrix reduced order model described in \cite{Tmatpaper}; and LSC (standing for Least Squares Collocation) referring to the approach used in \cite{ChapmanHewettTrefethen2015}.

To use the system \eqref{QL-A-sol} in practice, we need to truncate the four indices $m$, $n$, $p$ and $q$. In all the results of this article, we truncate the indices $m$ and $n$ at the same number, $N$ say, and truncate the indices $p$ and $q$ at a larger value $1.2N$ (rounded up to the nearest integer). The latter truncation is not an optimised value, though it was chosen through trial and error to best balance the accuracy of computing the final results while keeping the computational cost low.

In the context of this article, the MSIA method is similar to QLNN because it also uses the Wiener--Hopf technique. However, it considers a different rearrangement of \eqref{QL-SoEs} which is given by \eqref{QL-SoEs-MSIA} and this corresponds to considering rows or columns as coupled semi-infinite arrays. The resulting linear system is similar to \eqref{QL-A-sol} but the matrix $(\mathcal{I}-\mathcal{M})$ has identity block matrices along the diagonal and the entries in $A^{\RED{(I)}}$ are given by the solutions to associated semi-infinite array problems. Although truncating this system is done in the same way as QLNN, it has one less index (likely losing either $p$ or $q$).

The TMAT method aims to solve the following equation,
\begin{align}
\BF{A}=T\BF{F},
\end{align}
for the vector of scattering coefficients $\BF{A}$, whose entries $A_{mn}$ are ordered in the same way as described in Figure \ref{fig:QL-reshape_4d_to_2d}. Here, $T$ is the T-matrix which, for Dirichlet boundary conditions, is simply the identity matrix multiplied by the coefficient $-\frac{J_0(ka)}{H^{(1)}_0(ka)}$. Lastly, the entries of the vector $\BF{F}$ are ordered in the same way as $\BF{A}$ and are given by 
\begin{align}
F_{mn}=\lim_{\BF{r}\rightarrow\BF{R}_{mn}}\left(\Phi(\BF{r})-A_{mn}H^{(1)}_0(k|\BF{r}-\BF{R}_{mn}|)\right),
\end{align}
where $\Phi$ is the total field. These entries are currently unknown but can be written in terms of the entries of $\BF{A}$ when Foldy's approximation is applied. Then this matrix equation becomes identical to the system of equations \eqref{QL-SoEs} and is solved iteratively using GMRES. 

The LSC method aims to create and solve an over-determined matrix system of the form,
\begin{align}
\mathbb{M}\BF{A}=\BF{F},
\end{align}
for the vector of scattering coefficients $\BF{A}$, whose entries $A_{mn}$ are ordered in the same way as before. Here, the entries of $\mathbb{M}$ are given by $H^{(1)}_0(k|\BF{r}-\BF{R}_{mn}|)$ which are evaluated at multiple collocation points on each of the scatterer boundaries and the entries of $\BF{F}$ are boundary data on the collocation points (i.e. $-\Phi_{\textrm{I}}(\BF{r})$). Both systems of the TMAT and LSC methods are truncated by putting a limit on the number of scatterers. They also give very similar results so we will be using the LSC method for comparing our results in this article, although \citep{Tmatpaper} does feature a short comparison between the MSIA and TMAT methods for a special case of a point scatterer wedge.

\begin{figure}[ht]\centering
\includegraphics[width=0.9\textwidth]{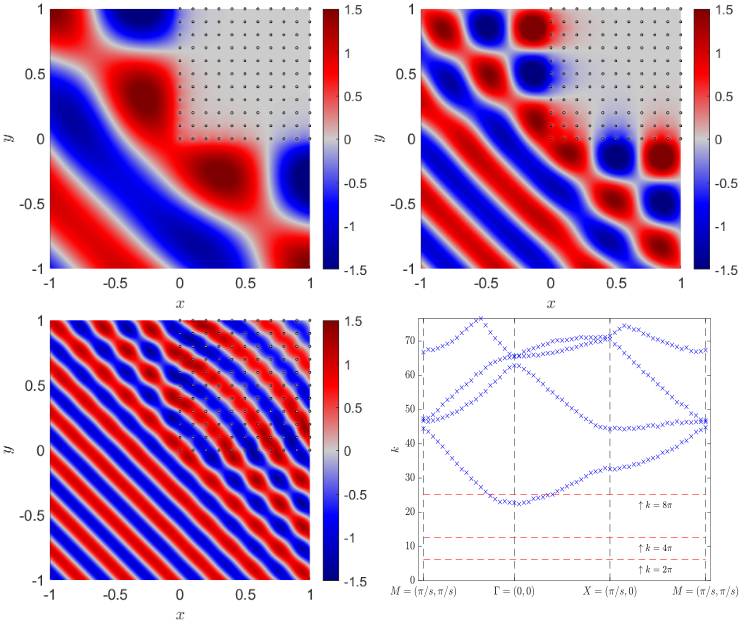}
\caption{Plots of the real part of the total field using the solution given by \eqref{QL-A-sol} and \eqref{QL-gensol} where we truncated the system at $N=100$. The plots on the top row and the bottom left have parameters, $s=0.1$, $a=0.001$ and $\theta_{\textrm{I}}=-\frac{3\pi}{4}$, with different wavenumbers, $k=2\pi$, $k=4\pi$ and $k=8\pi$ respectively. The plot on the bottom right is the dispersion diagram for a fully infinite doubly periodic set of scatterers with $s=0.1$ and $a=0.001$. The red dashed lines mark the wavenumbers that we considered for our test cases.}
\label{fig:QL_tf_results}
\end{figure}

Figure \ref{fig:QL_tf_results} plots the real part of the total field using the QLNN solution given by \eqref{QL-A-sol} where the system is truncated at $N=100$. The plots are displayed alongside the dispersion diagram for the fully infinite, doubly periodic case when the scatterers lie on the whole plane. In all cases, the same spacing $s=0.1$, scatterer size $a=0.001$ and incident angle $\theta_{\textrm{I}}=-\frac{3\pi}{4}$ have been used. The plots on the top left, top right and bottom left have different wavenumbers, $k=2\pi$, $k=4\pi$ and $k=8\pi$ respectively. 

When selecting the wavenumbers for these test cases, the intention was to try to test as wide a range as possible without pushing the limits of Foldy's approximation. The bottom right plot of Figure \ref{fig:QL_tf_results} displays the band gap diagram of a fully infinite doubly periodic set of scatterers with $s=0.1$ and $a=0.001$. This diagram was created by a MATLAB script that used the plane wave expansion method as described in \citep{EllisBarnwell-thesis}, which was also validated by another method in \citep{RuthVoiseyThesis}. The first of our chosen wavenumbers, $k=2\pi$ (implying $ka\approx0.0063$), is comfortably within the band gap, so the total field is expected to decay exponentially as it progresses deeper within the quarter lattice. The second wavenumber, $k=4\pi$ (implying $ka\approx0.0126$), is closer to the edge of the band gap so the decay here is still expected to decay exponentially, albeit at a slower rate. The last wavenumber $k=8\pi$ (implying $ka\approx0.0251$) lies within a pass band so that waves can propagate in the structure without attenuation (see for example \citep{TymisThompson2011}). All three test cases of Figure \ref{fig:QL_tf_results} display the expected behaviour within the quarter lattice. 

\begin{figure}[ht]\centering
\includegraphics[width=0.329\textwidth]{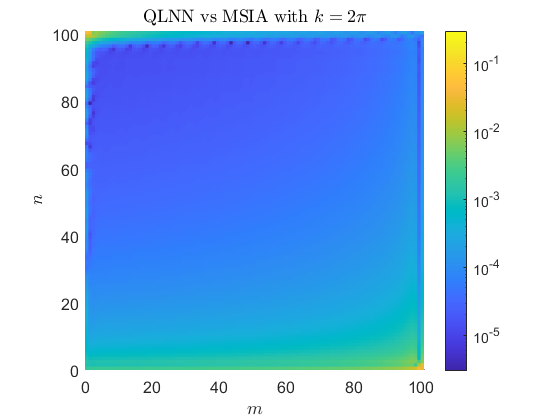}
\includegraphics[width=0.329\textwidth]{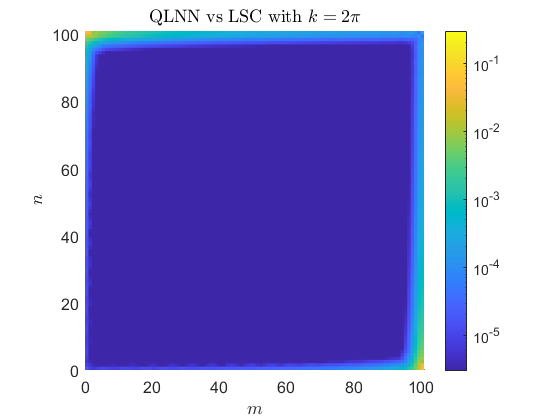}
\includegraphics[width=0.329\textwidth]{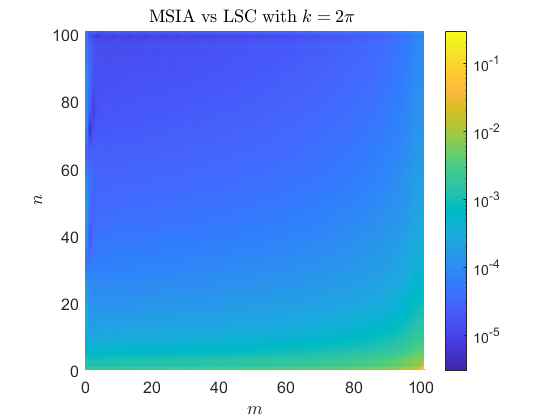}\\
\includegraphics[width=0.329\textwidth]{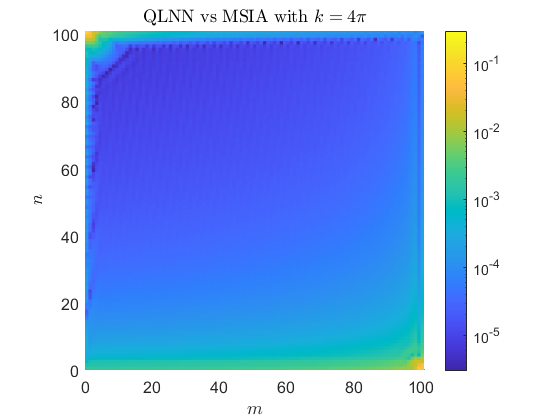}
\includegraphics[width=0.329\textwidth]{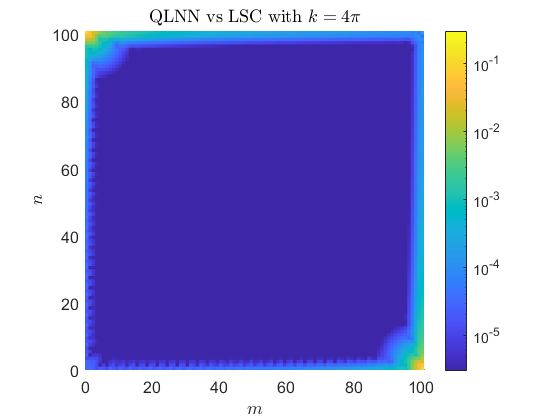}
\includegraphics[width=0.329\textwidth]{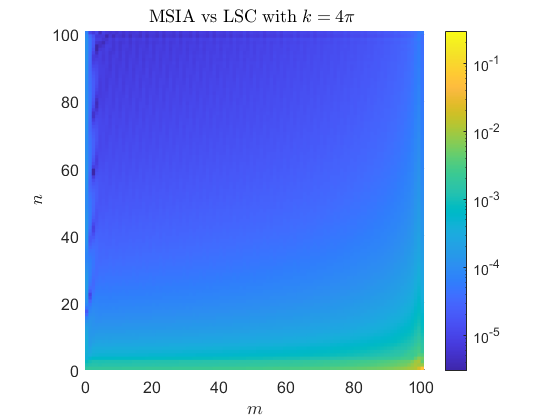}\\
\includegraphics[width=0.329\textwidth]{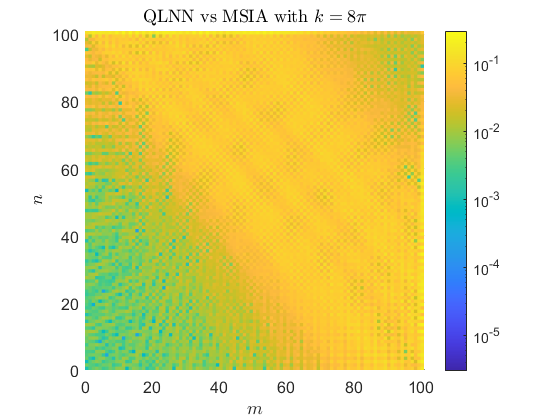}
\includegraphics[width=0.329\textwidth]{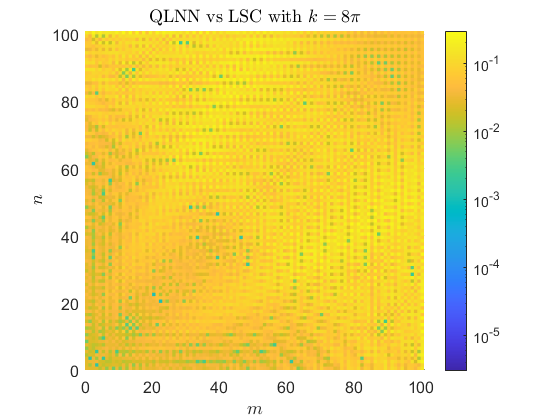}
\includegraphics[width=0.329\textwidth]{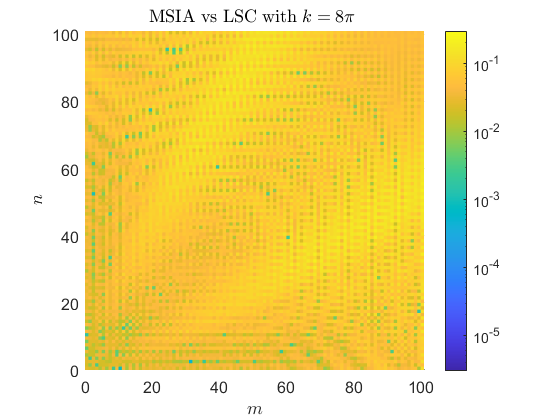}
\caption{Plots of the absolute error between different methods of working out the scattering coefficients $A_{mn}$ where we truncated at $N=100$. The left column plots the absolute error between the QLNN and MSIA methods, the middle column plots the absolute error between the QLNN and LSC methods, and the bottom column plots the absolute error between the MSIA and LSC methods. We also have the parameters, $s=0.1$, $a=0.001$ and $\theta_{\textrm{I}}=-\frac{3\pi}{4}$, with different wavenumbers, $k=2\pi$, $k=4\pi$ and $k=8\pi$ for the top, middle and bottom rows respectively.}
\label{fig:QL_Amn_error}
\end{figure}

In Figure \ref{fig:QL_Amn_error}, we consider the test cases presented in Figure \ref{fig:QL_tf_results} and plot the absolute difference of the scattering coefficients obtained by two of the three different methods. 
%The parameters are given by $s=0.1$, $a=0.001$ and $\theta_{\textrm{I}}=-\frac{3\pi}{4}$ with the top, middle and bottom rows looking at the test cases presented in the top left, bottom left and bottom left plots in Figure \ref{fig:QL_tf_results} (i.e. the cases with wavenumbers $k=2\pi$, $4\pi$ and $8\pi$) respectively with the truncation at $N=100$. 
The left column looks at the absolute difference, at each of the considered wavenumbers, between the QLNN and MSIA methods, the middle column looks at the absolute difference between the QLNN and LSC methods and the right column looks at the absolute difference between the MSIA and LSC methods. We find that the overall differences are smaller when $ks$ is smaller (band gap cases) and that the largest differences are closest to the truncated edges ($m=100$ or $n=100$) which will also improve as the truncation increases. We also find that in band gap cases, the QLNN and LSC solutions are virtually identical, even reaching machine precision in the centres of the relevant plots. Another feature to note is that because each of these cases are symmetric about the diagonal line from bottom left to top right, the difference between the QLNN and LSC solutions is symmetric along the same diagonal as well (barring some numerical error) whereas the plots involving the MSIA method are not. This is to be expected as the former two methods are better designed for modelling the interaction between neighbouring scatterers so are more likely to capture the symmetry. 

\begin{figure}[ht]\centering
\includegraphics[width=0.329\textwidth]{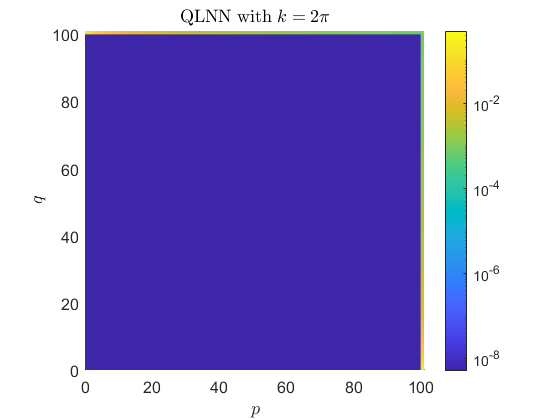}
\includegraphics[width=0.329\textwidth]{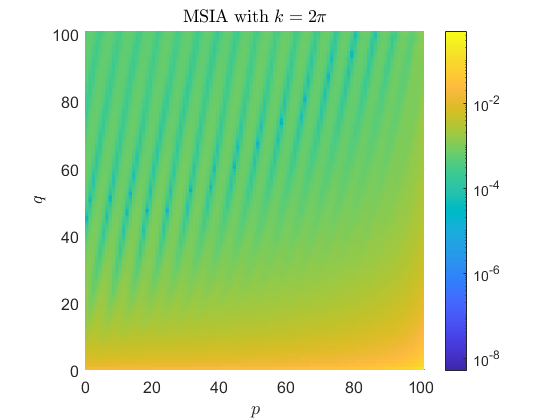}
\includegraphics[width=0.329\textwidth]{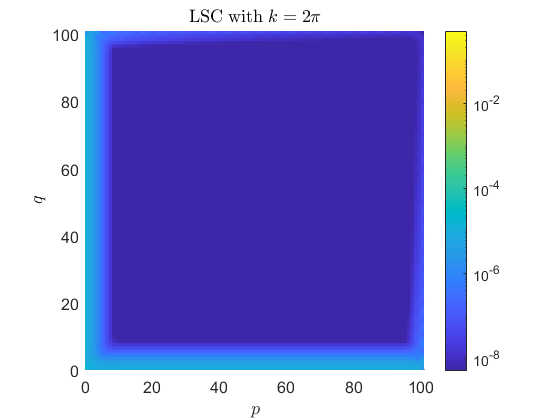}\\
\includegraphics[width=0.329\textwidth]{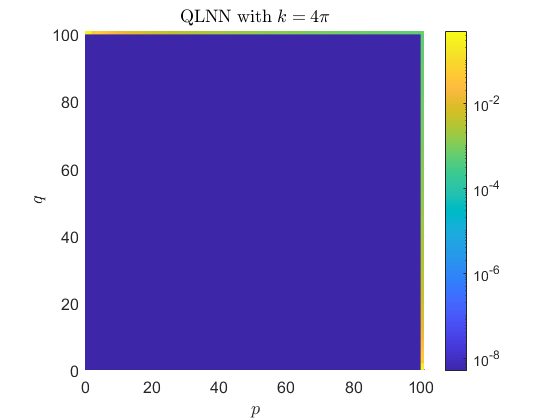}
\includegraphics[width=0.329\textwidth]{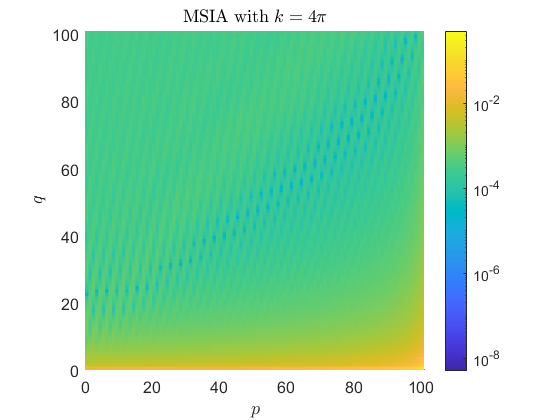}
\includegraphics[width=0.329\textwidth]{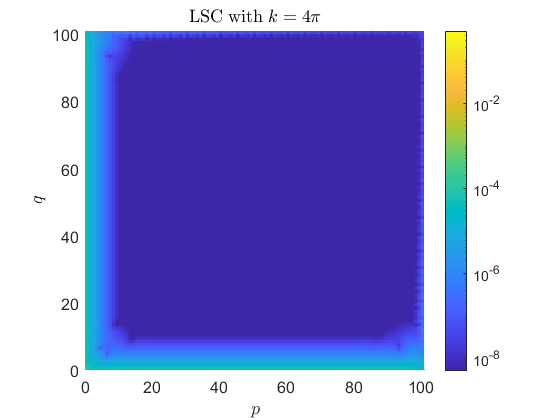}\\
\includegraphics[width=0.329\textwidth]{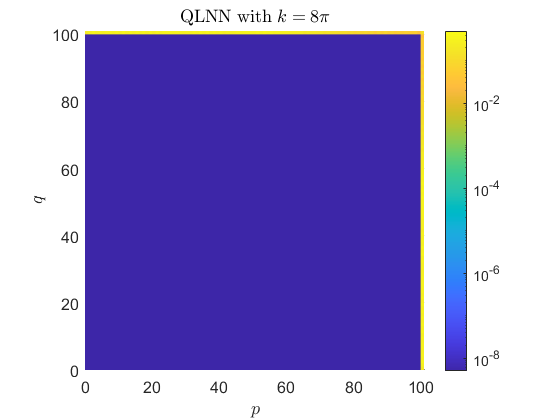}
\includegraphics[width=0.329\textwidth]{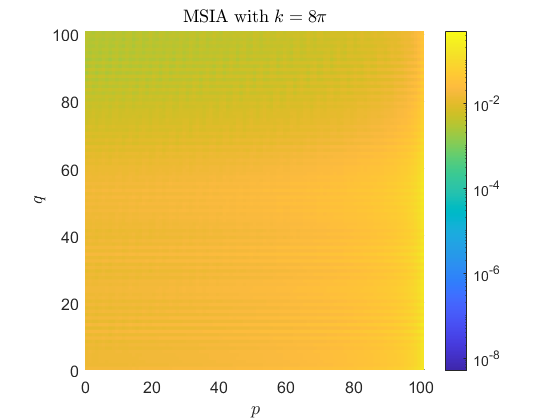}
\includegraphics[width=0.329\textwidth]{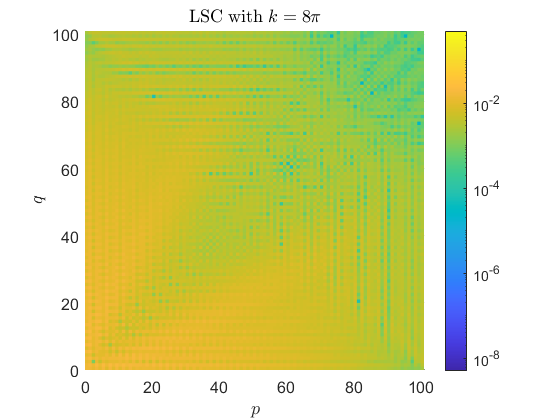}
\caption{Plots of the absolute error in the system of equations given by \eqref{QL-SoEs} where we have truncated the system at $N=100$ and substituted the different solutions for the scattering coefficients $A_{mn}$. The left, middle and right columns use the QLNN, MSIA and LSC solution respectively. We also have the parameters, $s=0.1$, $a=0.001$ and $\theta_{\textrm{I}}=-\frac{3\pi}{4}$, with different wavenumbers, $k=2\pi$, $k=4\pi$ and $k=8\pi$ for the top, middle and bottom rows respectively.}
\label{fig:QL_SoEs_error}
\end{figure}

Another way for us to compare the different methods is to substitute their respective solutions into the system of equations given by \eqref{QL-SoEs} (which is truncated at $N=100$) to visualise the resulting absolute error. Figure \ref{fig:QL_SoEs_error} displays this absolute error for all three methods with the same test cases as Figure \ref{fig:QL_tf_results}. 
%The parameters are given by $s=0.1$, $a=0.001$ and $\theta_{\textrm{I}}=-\frac{3\pi}{4}$ with the top, middle and bottom rows having wavenumbers $k=2\pi$, $k=4\pi$ and $k=8\pi$ respectively. The left, middle and right columns use the QLNN, MSIA and LSC solutions respectively. 
For the QLNN method, we find that the error in the systems of equations is virtually zero everywhere except on the truncated edges ($m=100$ or $n=100$) and that it overall satisfies the systems of equations best out of the three methods, the LSC method being second best. Another advantage of the QLNN method over the MSIA method is that it preserves the symmetry in symmetric test cases, as shown clearly (barring some numerical error) in the left test cases. This is due to the differences in how the quarter lattice is decomposed between the QLNN and MSIA methods (i.e. nearest neighbours vs rows).

\begin{figure}[ht]\centering
\includegraphics[width=0.329\textwidth]{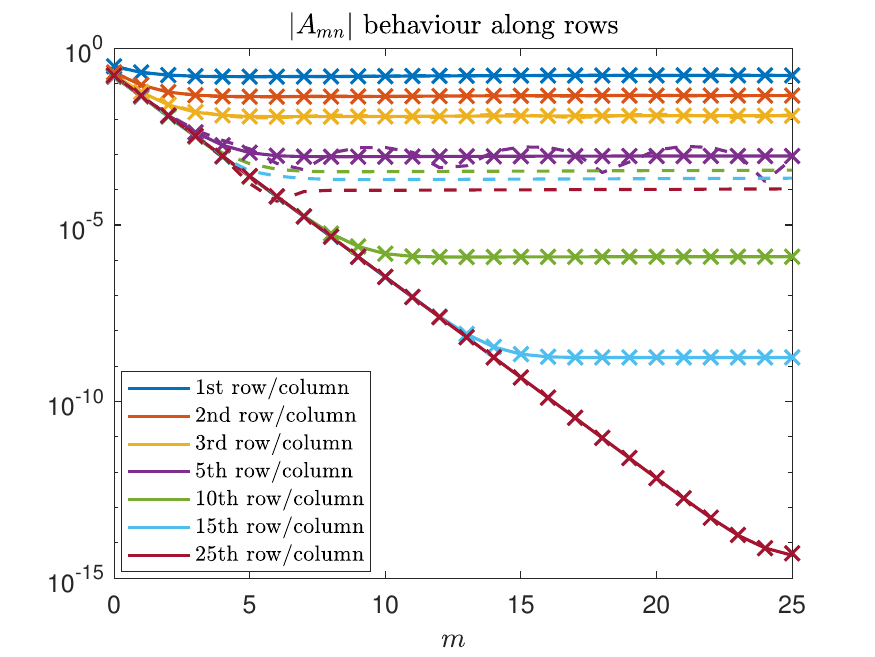}
\includegraphics[width=0.329\textwidth]{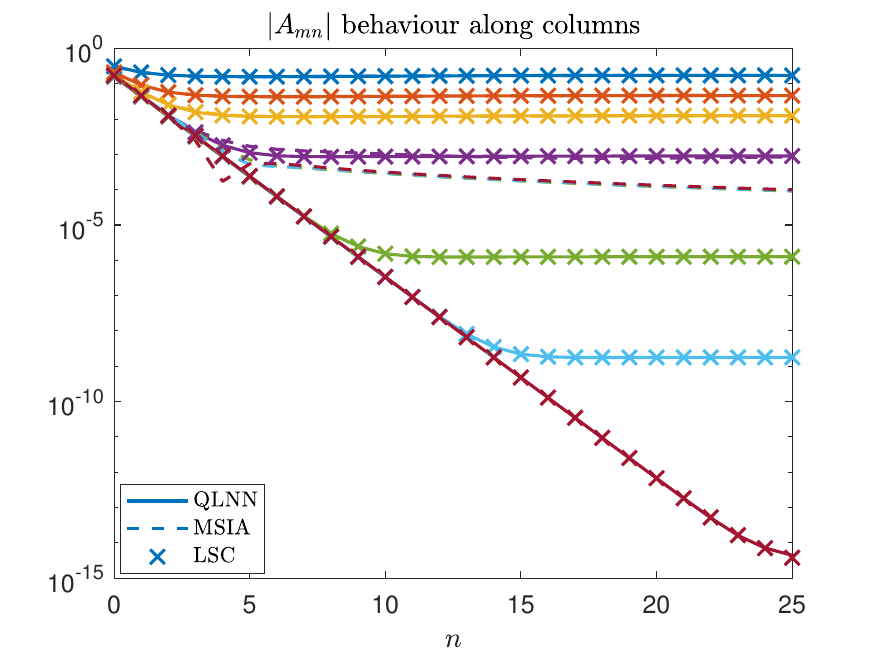}
\includegraphics[width=0.329\textwidth]{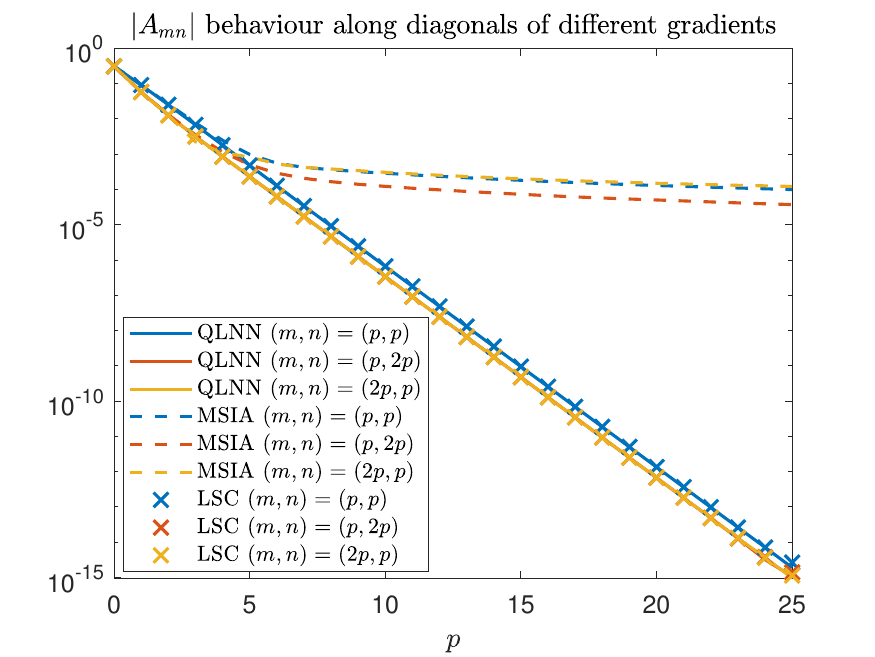}
\caption{Plots of $|A_{mn}|$ along various rows, columns or diagonals of the quarter lattice for the left,middle and right side respectively. For the left and middle plots, we display the first 25 values of $|A_{mn}|$ for the $1^{\textrm{st}}$, $2^{\textrm{nd}}$, $3^{\textrm{rd}}$, $5^{\textrm{th}}$, $10^{\textrm{th}}$, $15^{\textrm{th}}$ and  $25^{\textrm{th}}$ rows/columns (distinguished by colour on the left legend) and for each method (distinguished by plot style on the middle legend). For the right plot, we display the first 25 values of $|A_{mn}|$ along diagonals with different gradients which are given by the parametrisations $(m,n)=(p,p)$, $(m,n)=(p,2p)$ and $(m,n)=(2p,p)$ as $p$ increases. All these plots have parameters, $k=2\pi$, $s=0.1$, $a=0.001$ and $\theta_{\textrm{I}}=-\frac{3\pi}{4}$ and we truncated the system at $N=100$.}
\label{fig:QL_Amn_asymp_stop}
\end{figure}

In the next set of plots, we are interested to see if we can verify the expected behaviour of the scattering coefficients using this method which could be useful for estimating some kind of  effective wavenumbers of the quarter lattice. Figure \ref{fig:QL_Amn_asymp_stop} plots the absolute value of the scattering coefficients along a selection of rows, columns and diagonals. Here, we are using the same parameters as the top left plot of Figure \ref{fig:QL_tf_results} (i.e. $k=2\pi$) which corresponds to a band gap. For the row/column plots, we find that each of the graphs are exponentially decaying (because the coefficients are progressing deeper into the lattice) until they reach the diagonal of the quarter lattice when it starts flattening out (because the coefficients are no longer progressing deeper into the quarter lattice). While both the QLNN and LSC methods agree with their result and capture the expected behaviour, the MSIA method is only able to somewhat emulate it.
%According to \cite{TymisThompson2011} for the semi-infinite lattice, in the absence of any Bloch waves that could occur within a pass band, we should expect $|A_{mn}|$ to decay to zero as one moves deeper into the lattice. 

We also come to similar conclusions when we look at the behaviour of $|A_{mn}|$ along the diagonals as displayed in the right plot of Figure \ref{fig:QL_Amn_asymp_stop}. Here, we look at diagonals with different gradients which are given by the parametrisations $(m,n)=(p,p)$, $(m,n)=(p,2p)$ and $(m,n)=(2p,p)$ as $p$ increases. The result shows that there is non-stop exponential decay because we are always progressing deeper into the lattice as we progress along these diagonals. Through this we can also show that we will have exponentially decaying behaviour along all diagonals with strictly positive gradients.

\begin{figure}[ht!]\centering
\includegraphics[width=0.99\textwidth]{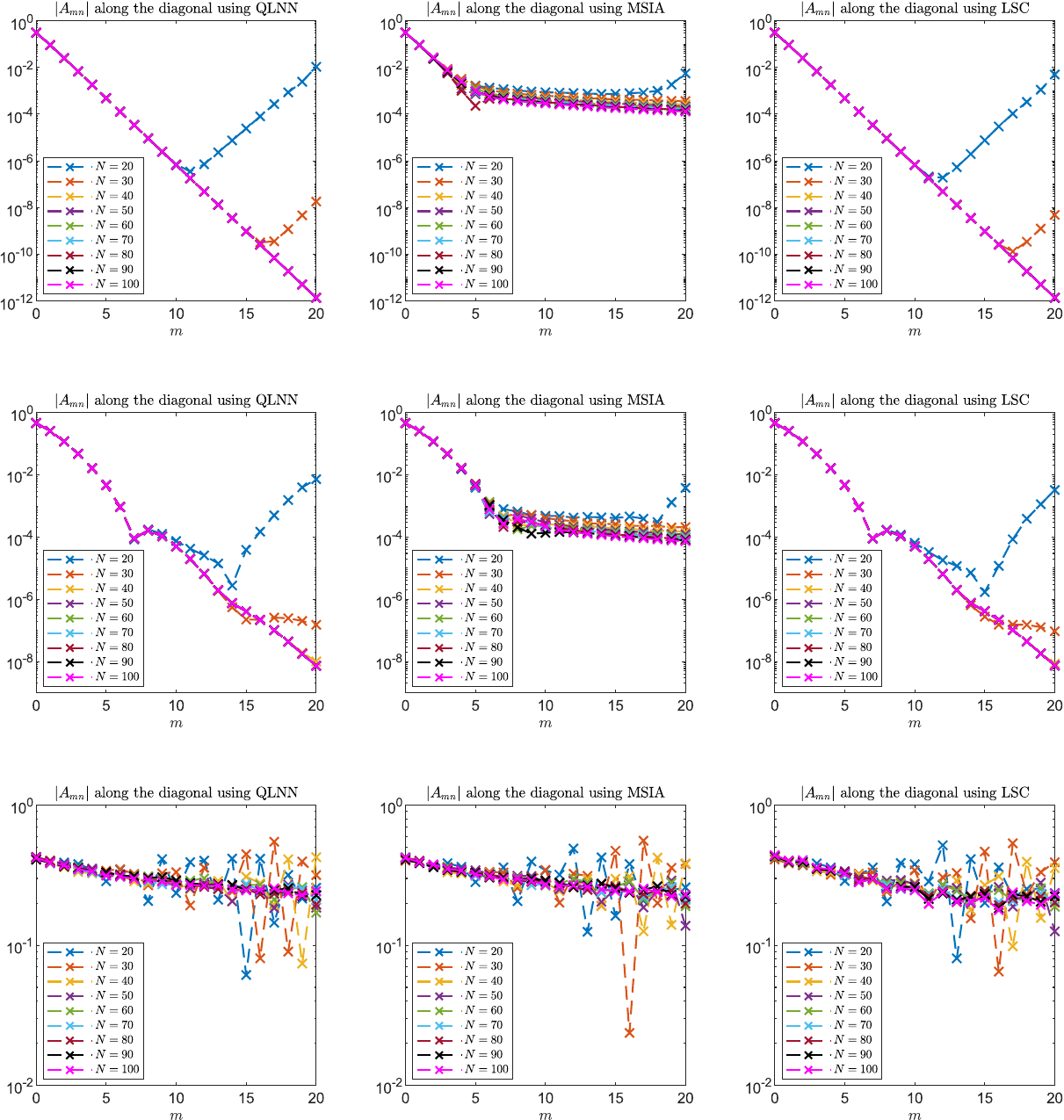}
\caption{Plots of $|A_{mn}|$ along the diagonal ($m=n$) with different truncations from $N=20$ to $100$. Here the left, middle and right columns use the QLNN, MSIA and LSC methods respectively. These plots also have parameters, $s=0.1$, $a=0.001$ and $\theta_{\textrm{I}}=-\frac{3\pi}{4}$ with the top, middle and bottom rows having wavenumbers $k=2\pi$, $4\pi$ and $8\pi$ respectively.}
\label{fig:QL_Amn_asymp_diag_bymethod}
\end{figure}

In Figure \ref{fig:QL_Amn_asymp_diag_bymethod}, we plot the first 20 diagonal values (i.e. $m=n$) of $|A_{mn}|$ for each of the test cases and each of the methods and vary the truncation. In each plot in this figure, we start with a smaller truncation $N=20$ and increase it in increments of 10 until we reach $N=100$. The top, middle and bottom rows have wavenumbers $k=2\pi$, $k=4\pi$ and $k=8\pi$ respectively. The left, middle and right columns use the QLNN, MSIA and LSC methods respectively. One noticeable observation (particularly of the $N=20$ and $N=30$ plots) in Figure \ref{fig:QL_Amn_asymp_diag_bymethod} is that at roughly $m=N/2$, the truncation is negating the expected behaviour of the scattering coefficients. Obviously, we can mitigate this error by increasing the truncation and we can see that by doing so, the rise in error occurs again at a further point but the original error has completely disappeared. This observation is not unique to the QLNN method and also occurs for the MSIA and LSC methods as well. Other observations are that as the wavenumber increases, the rate of decay in the band gap cases becomes slower, and that while the band gap cases clearly are converging as the truncation increases, the pass band case requires a much higher truncation in order to reach a small relative error. This is to be expected in the latter case, as we expect the interaction between scatterers (even those far away) to be much stronger.

Another main benefit to using either the QLNN or MSIA methods over more numerical approaches is that computing the matrix $\mathcal{M}$ in \eqref{QL-A-sol} is completely independent of the incident angle $\theta_{\textrm{I}}$, meaning that once $\mathcal{M}$ is determined and stored, it is fairly simple and quick to compute the scattering coefficients for a wide range of $\theta_{\textrm{I}}$ values with little computational cost. There may also be situations where one would want to try smaller truncations as a proof of concept and then opt for higher truncations for a more accurate result. While we have not implemented this in our scripts, it is certainly possible to efficiently compute the matrix $\mathcal{M}$ for higher truncations by taking the smaller precomputed $\mathcal{M}$ and augmenting it with the extra entries required to compute the more accurate solution.

\section{Conclusions}
In this article, we considered the problem of wave scattering by a doubly periodic quadrant of scatterers arranged in a square formation which we call the quarter lattice. We reduced it to a Wiener--Hopf equation in two complex variables with three unknown functions. Then by following the approach of \citep{Kisil2023}, we restricted ourselves to a manifold on which the two-dimensional Wiener--Hopf kernel is zero. Hence we were able to further reduce the problem to a Wiener--Hopf equation in one complex variable which can be solved explicitly (see \cite{Kisil2023} for details), and then find the sought-after scattering coefficients.

Through this new method, we are able to better model the nearest neighbour interactions between the scatterers as opposed to our previous method which would decompose the lattice into rows or columns. In the row approach the coupling between the rows was achieved via the forcing term and implemented numerically. In the new method due to the regions of interactions overlapping there is some analytical doubly periodic structure present in the solution. This new method is also more likely to preserve any symmetry in test cases where the incident field is symmetric with respect to the diagonal of the lattice.

Our method was critically compared to three other approaches for three meaningful test cases and it was found to perform at least as well as the others, sometimes performing better. From these comparisons, we concluded that the new method solves the truncated system of equations best of the three methods and the distribution of the error leans more towards the truncated edges which improves as the truncation increases. 

%We also managed to compare this new method with a previous one described in our article \citep{MSIApaper}, and a least squares collocation method which we have also used previously in said article. We did this by looking at the absolute difference between the resulting scattering coefficients and the absolute value of the error that occurs from substituting the coefficients into the system of equations \eqref{QL-SoEs} as well as looking at the behaviour of the coefficients along the rows, columns and diagonals of the lattice.

We show that all methods work better for smaller wavenumbers when there are less interactions between the scatterers, and we find that in cases of pass bands, all techniques require a larger truncation to sufficiently converge as shown in Figure \ref{fig:QL_Amn_asymp_diag_bymethod}. One may wonder if we could consider any of these methods to be the gold standard. In truth, a gold standard method is very difficult to derive because there will always be an approximation, assumption, simplification or truncation that is applied either in the process of finding a solution or when using it practically. For example, the Wiener--Hopf methods are technically able to find exact solutions for a full quadrant of scatterers, but to use practically, we need to introduce truncation. Whereas numerical methods are only able to find solutions for a finite amount of scatterers and would never be able to find a solution for a full quadrant of scatterers. Lastly, there are asymptotic methods as well (e.g. \citep{SchnitzerCraster2017}) which may do better in specific regimes. For example, approximating a quarter lattice as a fully infinite lattice would give accurate estimations for scattering coefficients deep within the lattice but will fail to accurately predict the corner behaviour. %a larger amount of scatterers but these are built as approximations from the start so may miss out important information in the higher order terms.

\section{Acknowledgements}

This research was supported by EPSRC grant EP/W018381/1. A.V.K. is supported by a Royal Society Dorothy Hodgkin Research Fellowship and a Dame Kathleen Ollerenshaw Fellowship.
Authors gratefully acknowledge the support of the EU H2020 grant MSCA-RISE-2020-101008140-EffectFact. The authors would also like to thank the Isaac Newton Institute for Mathematical Sciences (INI) for their support and hospitality during the programme ``Mathematical theory and applications of multiple wave scattering" (MWS) and ``WHT Follow on: the applications, generalisation and implementation of the Wiener--Hopf Method'' (WHTW02), where work on this paper was undertaken and supported by EPSRC grant no EP/R014604/1.

\bibliographystyle{abbrvnat}
\bibliography{WLbib}

\appendix
\section{Detailed derivation of Wiener--Hopf functional equation}\label{App:Func_eqn}
To recover \eqref{QL-AK-WHE}, we multiply \eqref{QL-AK-SoEs} by $z^{p+1}\zeta^{q+1}$ and sum over all $p,q\geq0$. Using the notations and \eqref{QL-AK-A++} introduced in Section \ref{Sec:Derive}, we can easily find that
\begin{align*}
\sum_{p,q=0}^{\infty} A_{pq}z^{p+1}\zeta^{q+1}=&z\zeta A^{++}(z,\zeta),\\
\sum_{p,q=0}^{\infty} e^{-ik\Theta(p, q)}z^{p+1}\zeta^{q+1}=&\frac{z\zeta}{(1-z_cz)(1-z_s\zeta)},\\
\sum_{p,q=0}^{\infty} \sum_{(m,n)\in\bar{S}_{pq}}^{\infty} A_{mn}\mathbb{H}_{(m-p)(n-q)}z^{p+1}\zeta^{q+1}=&\sum_{p,q=0}^{\infty}S_A(p,q)z^{p+1}\zeta^{q+1},
\end{align*}
therefore we obtain directly
\begin{align}\label{App:QL-WHE-with-T}
\gchoice z\zeta A^{++}(z,\zeta)+z\zeta\mathbb{T}=&z\zeta F^{++}(z,\zeta),
\end{align}
where $F^{++}$ is defined as in \eqref{QL-AK-WHE-forcing} and the remaining term $\mathbb{T}$ is given by
\begin{align}\label{App:QL-T}
\mathbb{T}=&\sum_{p,q=0}^{\infty}\sum_{\substack{m,n\in S_{pq}\\(m,n)\neq (p,q)}} A_{mn}\mathbb{H}_{(m-p)(n-q)}z^p\zeta^q.
\end{align}
Upon expanding the $(m,n)$ sum and noting that
\begin{align*}
\mathbb{H}_{(-1)(0)}=\mathbb{H}_{(1)(0)}=\mathbb{H}_{(0)(-1)}=\mathbb{H}_{(0)(1)}
=&H_0^{(1)}(ks),\\
\mathbb{H}_{(1)(1)}=\mathbb{H}_{(-1)(1)}=\mathbb{H}_{(1)(-1)}=\mathbb{H}_{(-1)(-1)}
=&H_0^{(1)}\left(ks\sqrt{2}\right),
\end{align*}
we can rewrite $\mathbb{T}$ as
\begin{align}\label{App:split-of-T}
\mathbb{T}=&H_0^{(1)}(ks)\mathbb{T}_1+H_0^{(1)}\left(ks\sqrt{2}\right)\mathbb{T}_2,
\end{align}
where
\begin{align}
\label{App:long-T1-new}\mathbb{T}_1=&\sum_{p,q=0}^{\infty}(A_{p-1,q}+A_{p+1,q}+A_{p,q-1}+A_{p,q+1})z^p\zeta^q,\\
\label{App:long-T2-new}\mathbb{T}_2=&\sum_{p,q=0}^{\infty}(A_{p-1,q-1}+A_{p-1,q+1}+A_{p+1,q-1}+A_{p+1,q+1})z^p\zeta^q.  
\end{align}
While noting that $A_{pq}=0$ whenever $p$ or $q$ are strictly negative, we use some basic algebra and the definitions in \eqref{QL-AK-A++} and \eqref{QL-AK-B+} to find the following expressions for the terms of $\mathbb{T}_1$,
\begin{align*}
\sum_{p,q=0}^{\infty}A_{p-1,q}z^p\zeta^q &=\sum_{p=1}^{\infty}\sum_{q=0}^{\infty}A_{p-1,q}z^p\zeta^q+\sum_{q=0}^{\infty}A_{-1,q}\zeta^q
=zA^{++}(z,\zeta),\\
\sum_{p,q=0}^{\infty}A_{p+1,q}z^p\zeta^q &=\sum_{p=-1}^{\infty}\sum_{q=0}^{\infty}A_{p+1,q}z^p\zeta^q-\frac{1}{z}\sum_{q=0}^{\infty}A_{0,q}\zeta^q
=\frac{1}{z}A^{++}(z,\zeta)-\frac{1}{z}B^{+}_2(\zeta),\\
\sum_{p,q=0}^{\infty}A_{p,q-1}z^p\zeta^q&=\zeta A^{++}(z,\zeta),\\
\sum_{p,q=0}^{\infty}A_{p,q+1}z^p\zeta^q&=\frac{1}{\zeta}A^{++}(z,\zeta)-\frac{1}{\zeta}B^{+}_1(z),
\end{align*}
which implies that
\begin{align}
\label{App:T1-parts}\mathbb{T}_1&=\left(z+\frac{1}{z}+\zeta+\frac{1}{\zeta}\right)A^{++}(z,\zeta)-\frac{1}{\zeta}B^{+}_1(z)-\frac{1}{z}B^{+}_2(\zeta).
\end{align}
Using the same techniques, we find the following expressions for the terms of $\mathbb{T}_2$,
\begin{align*}
\sum_{p,q=0}^{\infty}A_{p-1,q-1}z^p\zeta^q
&=\sum_{p=1}^{\infty}\sum_{q=1}^{\infty}A_{p-1,q-1}z^p\zeta^q+\sum_{p=1}^{\infty}A_{p-1,-1}z^p+\sum_{q=0}^{\infty}A_{-1,q-1}\zeta^q,\\
&=z\zeta A^{++}(z,\zeta),\\
\sum_{p,q=0}^{\infty}A_{p-1,q+1}z^p\zeta^q
&=\sum_{p=1}^{\infty}\sum_{q=-1}^{\infty}A_{p-1,q+1}z^p\zeta^q-\frac{1}{\zeta}\sum_{p=1}^{\infty}A_{p-1,0}z^p+\sum_{q=0}^{\infty}A_{-1,q+1}\zeta^q,\\
&=\frac{z}{\zeta}A^{++}(z,\zeta)-\frac{z}{\zeta}B^{+}_1(z),\\
\sum_{p,q=0}^{\infty}A_{p+1,q-1}z^p\zeta^q
&=\sum_{p=-1}^{\infty}\sum_{q=1}^{\infty}A_{p+1,q-1}z^p\zeta^q+\sum_{p=0}^{\infty}A_{p+1,-1}z^p-\frac{1}{z}\sum_{q=1}^{\infty}A_{0,q-1}\zeta^q,\\
&=\frac{\zeta}{z}A^{++}(z,\zeta)-\frac{\zeta}{z}B^{+}_2(z),\\
\sum_{p,q=0}^{\infty}A_{p+1,q+1}z^p\zeta^q
&=\sum_{p=-1}^{\infty}\sum_{q=-1}^{\infty}A_{p+1,q+1}z^p\zeta^q-\frac{1}{\zeta}\sum_{p=-1}^{\infty}A_{p+1,0}z^p-\frac{1}{z}\sum_{q=0}^{\infty}A_{0,q+1}\zeta^q,\\
&=\frac{1}{z\zeta}A^{++}(z,\zeta)-\frac{1}{z\zeta}B^{+}_1(z)-\frac{1}{z\zeta}B^{+}_2(\zeta)+\frac{A_{00}}{z\zeta},
\end{align*}
which implies that
\begin{align}
\mathbb{T}_2&=\left(z+\frac{1}{z}\right)\left(\zeta+\frac{1}{\zeta}\right)A^{++}(z,\zeta)-\frac{1}{\zeta}\left(z+\frac{1}{z}\right)B^{+}_1(z)
\label{App:T2-parts}-\frac{1}{z}\left(\zeta+\frac{1}{\zeta}\right)B^{+}_2(z)+\frac{A_{00}}{z\zeta}.
\end{align}

And then upon substituting \eqref{App:split-of-T}, \eqref{App:T1-parts} and \eqref{App:T2-parts} into \eqref{App:QL-WHE-with-T}, we obtain
\begin{align*}
z\zeta&\tilde{K}(z,\zeta)A^{++}(z,\zeta)-zL_2(z)B_1^+(z)-\zeta L_2(\zeta)B_2^+(\zeta)+A_{00}H_0^{(1)}\left(ks\sqrt{2}\right)=z\zeta F^{++}(z,\zeta),
\end{align*}
which is exactly the sought-after functional equation \eqref{QL-AK-WHE}, where $\tilde{K}(z,\zeta)$ and $L_2(z)$ are given by \eqref{QL-AK-kernelK} and \eqref{QL-AK-kernelL2} respectively.
%\begin{align*}
%K(z,\zeta)=&H_0^{(1)}(ka)+H_0^{(1)}(ks)\left(z+\frac{1}{z}+\zeta+\frac{1}{\zeta}\right)+H_0^{(1)}\left(ks\sqrt{2}\right)\left(\zeta+\frac{1}{\zeta}\right)\left(z+\frac{1}{z}\right),\\
%L_2(z)=&H_0^{(1)}(ks)+\left(z+\frac{1}{z}\right)H_0^{(1)}\left(ks\sqrt{2}\right).\\
%\end{align*}

\section{Asymptotics of the function $M(z)$}\label{App:asympM}
In this appendix, we use the notation $t_2$ to indicate the plus solution to the quadratic equation given by $z^2+t_1z+1=0$ where $t_1\in\CC$ (i.e. $z=t_2=-\frac{t_1}{2}+\sqrt{\frac{t_1}{2}-1}\sqrt{\frac{t_1}{2}+1}$ where the square roots take the principal branch). On this selected branch, the absolute value of the solution is bounded $|t_2|\leq1$, for all $t_1\in\CC$ so this solution is always inside the unit circle. The reason $|t_2|$ is bounded is because the formula $t_2=-\frac{t_1}{2}+\sqrt{\left(\frac{t_1}{2}\right)^2-1}$ is the inverse of a Joukowsky transformation given by $t_1=-t_2-\frac{1}{t_2}$ which maps the inside of the unit circle $|t_2|\leq1$ to the entire $t_1$ complex plane. Additionally, the other solution to the quadratic equation will be $1/t_2$ which is outside the unit circle.

%\begin{align*}
%M(z)^{\pm1}&=-\frac{L_1(z)}{2L_2(z)}\pm\sqrt{\left(\frac{L_1(z)}{2L_2(z)}\right)^2-1},\\
%L_1(z)&=H^{(1)}_0(ka)+\left(z+\frac{1}{z}\right)H^{(1)}_0\left(ks\right),\\ 
%L_2(z)&=H^{(1)}_0(ks)+\left(z+\frac{1}{z}\right)H^{(1)}_0\left(ks\sqrt{2}\right),
%\end{align*}
Recall the formula for $M(z)$ given by \eqref{QL-AK-manifold-def}. Let $a_1=\frac{\gchoice}{H^{(1)}_0(ks)}$, $b_1=\frac{H^{(1)}_0(ks)}{H^{(1)}_0(ks\sqrt{2})}$ for simplicity and then 
$$\frac{L_1(z)}{L_2(z)}=\frac{b_1(z^2+a_1z+1)}{z^2+b_1z+1}=b_1+\frac{b_1(a_1-b_1)z}{z^2+b_1z+1}.$$
The only potential pole locations of $M(z)$ are solutions of the quadratic $z^2+b_1z+1=0$ called $b_2$ and $1/b_2$ (located inside and outside the unit circle respectively), which are also the zeros of $L_2(z)$. Let $z\rightarrow b_2$ in $\frac{L_1(z)}{L_2(z)}$, then we find that
$$\frac{L_1(z)}{L_2(z)}=\frac{(a_1-b_1)b_1b_2^2}{(b_2^2-1)(z-b_2)}+\left(1-\frac{(a_1-b_1)b_1b_2}{(b_2^2-1)^2}\right) +\frac{(a_1-b_1)b_1b_2^2}{(b_2^2-1)^3}(z-b_2)+O\left((z-b_2)^2\right)$$
%For ease of notation, we rewrite the above,
%$$\frac{L_1(z)}{L_2(z)}=\frac{A}{z-b_2}+B+C(z-b_2)+O\left((z-b_2)^2\right),$$
%then we have
%$$\sqrt{\left(\frac{L_1(z)}{2L_2(z)}\right)^2-1}%=\sqrt{\frac{A^2}{4(z-b_2)^2}+\frac{AB}{2(z-b_2)}+\frac{B^2+2AC}{4}-1+O\left((z-b_2)\right)}$$
%=\frac{A}{2(z-b_2)}+\frac{B}{2}+\frac{AC-2}{2A}(z-b_2)+O\left((z-b_2)^2\right),$$
After substituting this asympotic expansion into $M(z)$, the first two terms in the resulting expansion cancel out. Hence, we obtain
\begin{align}
M(z)=-\frac{(b_2^2-1)}{(a_1-b_1)b_1b_2^2}(z-b_2)+O\left((z-b_2)^2\right).
\end{align}
This means that the pole at $z=b_2$ is actually a removable singularity and the local behaviour is that of a simple zero instead. Similarly if $z\rightarrow1/b_2$ then
\begin{align}
M(z)=-\frac{(1-b_2^2)}{(a_1-b_1)b_1}\left(z-\frac{1}{b_2}\right)+O\left(\left(z-\frac{1}{b_2}\right)^2\right).
\end{align}

When we substitute $\frac{L_1(z)}{L_2(z)}$ into $M(z)$, it can be rewritten in the following form
$$M(z)=-\frac{b_1(z^2+a_1z+1)}{2(z^2+b_1z+1)}+\frac{1}{2}\sqrt{\frac{(b_1^2-4)(z^2+c_1z+1)(z^2+d_1z+1)}{(z^2+b_1z+1)^2}},$$
where $c_1=\frac{b_1(a_1-2)}{b_1-2}$ and $d_1=\frac{b_1(a_1+2)}{b_1+2}$. This means that the roots of the quadratic equations $z^2+c_1z+1=0$ and $z^2+d_1z+1=0$ are the locations of the branch points of $M(z)$, called $c_2$, $1/c_2$, $d_2$ and $1/d_2$. Here, the branch points $c_2$ and $d_2$ are inside the unit circle and are the start and end points of one branch cut. The other branch points $1/c_2$ and $1/d_2$ are outside the unit circle and are the start and end points of the other branch cut.
%$$M(z)^{\pm1}=-\frac{b_1(z-a_2)(z-1/a_2)}{2(z-b_2)(z-1/b_2)}\pm\frac{\sqrt{b_1^2-4}}{2(z-b_2)(z-1/b_2)}\sqrt{(z-c_2)}\sqrt{(z-d_2)}\sqrt{(z-1/c_2)}\sqrt{(z-1/d_2)}.$$

The asymptotic behaviour of $M(z)$ as $z\rightarrow c_2$ is given by,
%$$M(z)^{\pm1}=-\frac{b_1(c_1-a_1)}{2(c_1-b_1)}\pm\frac{\sqrt{b_1^2-4}}{2(b_1-c_1)c_2}\sqrt{c_2-\frac{1}{c_2}}\sqrt{c_2-d_2}\sqrt{c_2-\frac{1}{d_2}}\sqrt{z-c_2}+O\left(z-c_2\right)$$
%$$M(z)^{\pm1}=-1\pm\frac{1}{2}\left(\frac{(b_1^2-4)\sqrt{c_1^2-4}(c_2-d_2)(c_2-1/d_2)(z-c_2)}{(b_1-c_1)^2c_2^2}\right)^{\frac{1}{2}}+O\left(z-c_2\right)$$
%$$M(z)^{\pm1}=-1\pm\left(\frac{(b_1-2)^3(b_1+2)\sqrt{c_1^2-4}(d_1-c_1)\left(\frac{z}{c_2}-1\right)}{4b_1^2(b_1-a_1)^2}\right)^{\frac{1}{2}}+O\left(z-c_2\right)$$
\begin{align}
M(z)=-1+\left(\frac{(b_1-2)^2\sqrt{c_1^2-4}\left(\frac{z}{c_2}-1\right)}{b_1(b_1-a_1)}\right)^{\frac{1}{2}}+O\left(\frac{z}{c_2}-1\right),
\end{align}
and the asymptotic behaviour of $M(z)$ as $z\rightarrow d_2$ is similarly given by,
%$$M(z)^{\pm1}=-\frac{b_1(d_1-a_1)}{2(d_1-b_1)}\pm\frac{\sqrt{b_1^2-4}}{2(b_1-d_1)d_2}\sqrt{d_2-\frac{1}{d_2}}\sqrt{d_2-c_2}\sqrt{d_2-\frac{1}{c_2}}\sqrt{z-d_2}+O\left(z-d_2\right)$$
%$$M(z)^{\pm1}=1\pm\left(\frac{(b_1^2-4)\sqrt{d_1^2-4}(c_1-d_1)(\frac{z}{d_2}-1)}{4(b_1-d_1)^2}\right)^{\frac{1}{2}}+O\left(z-d_2\right)$$
\begin{align}
M(z)=1+\left(\frac{(b_1+2)^2\sqrt{d_1^2-4}\left(\frac{z}{d_2}-1\right)}{b_1(a_1-b_1)}\right)^{\frac{1}{2}}+O\left(\frac{z}{d_2}-1\right).
\end{align}
After noting that $M(z)$ is its own inverse function, we find that the branch point locations can be written in terms of itself, i.e. $c_2=M(-1)$ and $d_2=M(1)$.

\section{Some useful integral evaluations}\label{App:integral}
In this appendix, we will give some more detail on how we derive the three integral evaluations given by \eqref{QL-A++-int123}, \eqref{QL-A++-int4-part1} and \eqref{QL-A++-int4-part2}. In what follows, we need to write the kernel in terms of the manifold function,
\begin{align}\label{App:integral-kernel}
K(z,\zeta)=\frac{L_2(z)}{\zeta}(\zeta-M(z))(\zeta-M(z)^{-1}),
\end{align}
and apply it to all three integrals. These integrals have a few similarities as well. They are all found to be a sum of residues of poles inside the integration contour $C_\zeta$. Two of these poles are simple ones at $\zeta=M(z)$ and $M(z)^{-1}$ which is obvious after applying \eqref{App:integral-kernel}, but only the $\zeta=M(z)$ pole is inside $C_\zeta$. They also all have a potential pole at $\zeta=0$ but with different orders.

After using \eqref{App:integral-kernel}, the first integral \eqref{QL-A++-int123} is simply the sum of the residues of the poles at $\zeta=M(z)$ and $\zeta=0$,
\begin{align}
\nonumber\frac{1}{2\pi i}\int_{C_\zeta}\frac{\zeta^{-n-2}}{K(z,\zeta)}\textrm{d}\zeta&=\frac{1}{2\pi i}\int_{C_\zeta}\frac{\zeta^{-n-1}}{L_2(z)(\zeta-M(z))(\zeta-M(z)^{-1})}\textrm{d}\zeta\\
\nonumber&=\underset{\zeta=M(z)}{\text{Res}}\left(\frac{\zeta^{-n-2}}{K(z,\zeta)}\right)+\underset{\zeta=0}{\text{Res}}\left(\frac{\zeta^{-n-2}}{K(z,\zeta)}\right).
\end{align}
Both of these residues can be determined explicitly using standard methods. The residue of the simple pole at $\zeta=M(z)$ is given by,
\begin{align}\label{App:integral-int1-resM}
\underset{\zeta=M(z)}{\text{Res}}\left(\frac{\zeta^{-n-2}}{K(z,\zeta)}\right)=\frac{M(z)^{-n}}{L_2(z)(M(z)^2-1)}.
\end{align}
The second pole at $\zeta=0$ is of order $n+1$ and after using partial fraction expansions, we find that the residue is given by,
\begin{align}\label{App:integral-int1-res0}
\nonumber\underset{\zeta=0}{\text{Res}}\left(\frac{\zeta^{-n-2}}{K(z,\zeta)}\right)
&=\frac{1}{n!}\left.\ParDer{^n}{\zeta^n}\left(\frac{1}{L_2(z)(\zeta-M(z))(\zeta-M(z)^{-1})}\right)\right|_{\zeta=0},\\
\nonumber&=\frac{1}{n!}\frac{M(z)}{L_2(z)(M(z)^2-1)}\left.\ParDer{^n}{\zeta^n}\left(\frac{1}{\zeta-M(z)}-\frac{1}{\zeta-M(z)^{-1}}\right)\right|_{\zeta=0},\\
%\nonumber&\tmcolor{magenta}{=\frac{1}{n!}\frac{M(z)}{L_2(z)(M(z)^2-1)}\left.\left(\frac{(-1)^nn!}{(\zeta-M(z))^{n+1}}-\frac{(-1)^nn!}{(\zeta-M(z)^{-1})^{n+1}}\right)\right|_{\zeta=0}},\\
%\nonumber&\tmcolor{magenta}{=\frac{M(z)}{L_2(z)(M(z)^2-1)}\left(M(z)^{n+1}-M(z)^{-n-1}\right),}\\
&=\frac{M(z)^{n+2}-M(z)^{-n}}{L_2(z)(M(z)^2-1)}.
\end{align}
Therefore, we add \eqref{App:integral-int1-resM} and \eqref{App:integral-int1-res0} to obtain the result,
\begin{align}\label{App:integral-int1}
\frac{1}{2\pi i}\int_{C_\zeta}\frac{\zeta^{-n-2}}{K(z,\zeta)}\textrm{d}\zeta&=\frac{M(z)^{n+2}}{L_2(z)(M(z)^2-1)}.
\end{align}

For the second integral \eqref{QL-A++-int4-part1}, we find that it is the sum of the residues of the simple pole at $\zeta=M(z)$ and the pole of order $n$ at $\zeta=0$. There is an additional pole at $\zeta=1/z_s$ but this is outside $C_\zeta$ so is not included. This means that,
\begin{align}
\nonumber\frac{1}{2\pi i}\int_{C_\zeta}\frac{\zeta^{-n-1}}{K(z,\zeta)(z_s\zeta-1)}\textrm{d}\zeta
&=\frac{1}{2\pi i}\int_{C_\zeta}\frac{\zeta^{-n}}{L_2(z)(\zeta-M(z))(\zeta-M(z)^{-1})(z_s\zeta-1)}\textrm{d}\zeta,\\
\nonumber &=\underset{\zeta=M(z)}{\text{Res}}\left(\frac{\zeta^{-n-1}}{K(z,\zeta)(z_s\zeta-1)}\right)
+\underset{\zeta=0}{\text{Res}}\left(\frac{\zeta^{-n-1}}{K(z,\zeta)(z_s\zeta-1)}\right).
\end{align}
Here, the residue for $\zeta=M(z)$ is given by 
\begin{align}\label{App:integral-int2-resM}
\underset{\zeta=M(z)}{\text{Res}}\left(\frac{\zeta^{-n-1}}{K(z,\zeta)(z_s\zeta-1)}\right)=\frac{M(z)^{1-n}}{L_2(z)(M(z)^2-1)(z_sM(z)-1)},
\end{align}
and the residue for $\zeta=0$ is given by 
\begin{align}\label{App:integral-int2-res0}
\nonumber\underset{\zeta=0}{\text{Res}}\left(\frac{\zeta^{-n-1}}{K(z,\zeta)(z_s\zeta-1)}\right)
=&\ \frac{1}{(n-1)!}\left.\ParDer{^{n-1}}{\zeta^{n-1}}\left(\frac{1}{L_2(z)(\zeta-M(z))(\zeta-M(z)^{-1})(z_s\zeta-1)}\right)\right|_{\zeta=0},\\
%\nonumber=&\tmcolor{magenta}{\ \frac{1}{(n-1)!L_2(z)}\ParDer{^{n-1}}{\zeta^{n-1}}\bigg(\frac{1}{(M(z)-M(z)^{-1})(z_sM(z)-1)(\zeta-M(z))}}\\
%\nonumber&\tmcolor{magenta}{+\frac{M(z)}{(M(z)-M(z)^{-1})(M(z)-z_s)(\zeta-M(z)^{-1})}-\frac{z_s^2M(z)}{(z_sM(z)-1)(M(z)-z_s)(z_s\zeta-1)}\bigg)\bigg|_{\zeta=0},}\\
%\nonumber=&\tmcolor{magenta}{\ -\frac{1}{L_2(z)}\bigg(\frac{M(z)^{-n}}{(M(z)-M(z)^{-1})(z_sM(z)-1)}}\\
%\nonumber&\tmcolor{magenta}{+\frac{M(z)^{n+1}}{(M(z)-M(z)^{-1})(M(z)-z_s)}-\frac{z_s^{n+1}M(z)}{(z_sM(z)-1)(M(z)-z_s)}\bigg),}\\
\nonumber=&\ -\frac{M(z)}{L_2(z)}\bigg[\frac{M(z)^{-n}}{(M(z)^2-1)(z_sM(z)-1)}\\
&+\frac{1}{(M(z)-z_s)}\left(\frac{M(z)^{n+1}}{M(z)^2-1}-\frac{z_s^{n+1}}{(z_sM(z)-1)}\right)\bigg],
\end{align}
where we used partial fraction expansions to obtain the result.
%$$\frac{1}{(\zeta-M(z))(\zeta-M(z)^{-1})(z_s\zeta-1)}=\frac{A}{(\zeta-M(z))}+\frac{B}{(\zeta-M(z)^{-1})}+\frac{C}{(z_s\zeta-1)}$$
%$$1=A(\zeta-M(z)^{-1})(z_s\zeta-1)+B(\zeta-M(z))(z_s\zeta-1)+C(\zeta-M(z))(\zeta-M(z)^{-1})$$
%$$\zeta=M(z)\implies A=\frac{1}{(M(z)-M(z)^{-1})(z_sM(z)-1)}$$
%$$\zeta=M(z)^{-1}\implies B=\frac{M(z)}{(M(z)-M(z)^{-1})(M(z)-z_s)}$$
%$$\zeta=1/z_s\implies C=-\frac{z_s^2M(z)}{(z_sM(z)-1)(M(z)-z_s)}$$
We then add \eqref{App:integral-int2-resM} and \eqref{App:integral-int2-res0} to recover \eqref{QL-A++-int4-part1},
\begin{align}\label{App:integral-int2}
\frac{1}{2\pi i}\int_{C_\zeta}\frac{\zeta^{-n-1}}{K(z,\zeta)(z_s\zeta-1)}\textrm{d}\zeta
&=-\frac{M(z)}{L_2(z)(M(z)-z_s)}\left[\frac{M(z)^{n+1}}{M(z)^2-1}-\frac{z_s^{n+1}}{(z_sM(z)-1)}\right].
\end{align}

For the last integral \eqref{QL-A++-int4-part2}, the evaluation is once again a sum of residues of the simple pole at $\zeta=M(z)$ plus a potential pole of order $n-q$ at $\zeta=0$, which is only present if $n>q$,
\begin{align}%\label{QL-A++-int4-part2}
\nonumber\frac{1}{2\pi i}\int_{C_\zeta}\frac{\zeta^{q-n-1}}{K(z,\zeta)}\textrm{d}\zeta
\nonumber&=\frac{1}{2\pi i}\int_{C_\zeta}\frac{\zeta^{q-n}}{L_2(z)(\zeta-M(z))(\zeta-M(z)^{-1})}\textrm{d}\zeta\\
\nonumber&=\underset{\zeta=M(z)}{\text{Res}}\left(\frac{\zeta^{q-n-1}}{K(z,\zeta)}\right)
+\mathcal{H}(n-q)\underset{\zeta=0}{\text{Res}}\left(\frac{\zeta^{q-n-1}}{K(z,\zeta)}\right),
\end{align}
where $\mathcal{H}(z)$ is the Heaviside function. Here, the residue for $\zeta=M(z)$ is given by 
\begin{align}\label{App:integral-int3-resM}
\underset{\zeta=M(z)}{\text{Res}}\left(\frac{\zeta^{q-n-1}}{K(z,\zeta)}\right)=\frac{M(z)^{q-n+1}}{L_2(z)(M(z)^2-1)},
\end{align}
and the residue for $\zeta=0$ is and given by 
\begin{align}\label{App:integral-int3-res0}
\underset{\zeta=0}{\text{Res}}\left(\frac{\zeta^{q-n-1}}{K(z,\zeta)}\right)
\nonumber=&\ \frac{1}{(n-q-1)!}\left.\ParDer{^{n-q-1}}{\zeta^{n-q-1}}\left(\frac{1}{L_2(z)(\zeta-M(z))(\zeta-M(z)^{-1})}\right)\right|_{\zeta=0},\\
%\nonumber=&\tmcolor{magenta}{\ \frac{M(z)}{(n-q-1)!L_2(z)(M(z)^2-1)}\left.\ParDer{^{n-q-1}}{\zeta^{n-q-1}}\left(\frac{1}{(\zeta-M(z))}-\frac{1}{(\zeta-M(z)^{-1})}\right)\right|_{\zeta=0},}\\
=&\ \frac{M(z)}{L_2(z)(M(z)^2-1)}\left(M(z)^{n-q}-M(z)^{q-n}\right).
\end{align}
Upon adding \eqref{App:integral-int3-resM} and \eqref{App:integral-int3-res0} together, we get the final result recovering \eqref{QL-A++-int4-part2}, 
\begin{align}\label{App:integral-int3}
\frac{1}{2\pi i}\int_{C_\zeta}\frac{\zeta^{q-n-1}}{K(z,\zeta)}\textrm{d}\zeta
%\nonumber&=\frac{M(z)^{q-n+1}}{L_2(z)(M(z)^2-1)}+\mathcal{H}(n-q)\frac{M(z)^{n-q+1}-M(z)^{q-n+1}}{L_2(z)(M(z)^2-1)}\\
&=\frac{M(z)^{|n-q|+1}}{L_2(z)(M(z)^2-1)}.
\end{align}

\end{document}